# Coherent perfect absorption of quantum light

Anton N. Vetlugin

*Centre for Disruptive Photonic Technologies, SPMS, TPI, Nanyang Technological University, Singapore 637371*

*a.vetlugin@ntu.edu.sg*

Coherent perfect absorption (CPA) was introduced as a classical optics phenomenon of a standing-wave absorption by a subwavelength film. In this paper, we develop a theory of CPA of quantized standing waves taking account of a subwavelength thickness of the absorber. This approach allows us to merge all known quantum effects of CPA under a single theory and introduce other regimes of CPA including CPA of NOON states with arbitrary number of entangled photons, orthogonally squeezed vacuum states, continuous variable entangled states, and Schrödinger cat states. Detailed analysis of the quantum regime of CPA sheds light on fundamental aspects of quantum light dissipation and its practical implementation in quantum optics and quantum information.

## 1. Introduction

Coherent perfect absorption (CPA) is a time-reversed process of a lasing where full absorption of classical light is achieved by coherent illumination of an absorber [1-3]. Importantly, an absorber of a subwavelength thickness can operate between the regimes of total absorption and total transmission [4]. Explanation of this phenomenon is given in a standing wave picture: if absorber is placed at anti-node of the standing wave, enhanced dissipation of light takes place, and if absorber is placed at node of the standing wave, no interaction happens since total electric field is equal to zero. Light-by-light control [4], all-optical switching [5], signal modulation [6,7], dark-pulse generation [8] and coherent amplifier [9] are clearly explained in this picture.

    In contrast, quantum regime of CPA, where thin absorber is coherently illuminated by quantum light, lacks this clarity of explanation. The outcome of CPA process strongly depends on the quantum state of light. For instance, 'classical' modulation between the regimes of total absorption and total transmission takes place for a single photon state [10,11] while probabilistic zero- or two-photon absorption may take place for two photon states [12-14]. Developed theoretical models of CPA of quantum light [15-17] describe the problem in terms of quantized travelling waves, Fig. 1(a), where subwavelength thickness of the absorber is not taken into account. Moreover, either bosonic [15] or fermionic [13] second quantization formalism is required depending on the quantum states under consideration. Despite the lack of clear picture of the underlying processes, quantum regime of CPA is of great interest for applications in quantum optics and quantum information. CPA provides a robust approach for quantum states control including quantum states filtering [16-18], manipulation of quantum light correlations [12-15,19] and implementation of the anti-Hong-Ou-Mandel effect [20]. Recently, the mechanism of distributed CPA of quantum light was proposed for deterministic generation of entanglement in multi-nodal quantum networks [21]. From fundamental point of view, quantum regime of CPA provides new insights on the process of quantum light absorption including local [10,11,22] and non-local [23] photon absorption control, probabilistic two-photon and deterministic one-photon absorption of two-photon states [12,14,20]. Further development of this research field requires clear explanation of quantum effects of CPA.

    In this paper, we develop theory of CPA describing CPA of quantum light in a unified and simple manner. We consider the problem in a proper basis of quantized standing waves, Fig. 1(b) and 1(c), taking account of a subwavelength thickness of the absorber. Our approach allows us to to merge all known quantum effects of CPA under a single theory and introduce other regimes of CPA including CPA of NOON states with arbitrary number of entangled photons, orthogonally squeezed vacuum, continuous variable entangled, Schrödinger cat states.

Developed here theory reveals a common nature between classical and quantum regimes of CPA. To emphasize similarities of these two regimes of CPA, we describe both effects in a parallel way. First, we describe the classical regime of CPA in the form where two travelling waves are considered as a superposition of two standing waves (Section 2). Next, in a similar way, we build a general description of CPA of quantum light (Section 3). Finally, we apply our approach to CPA of discrete (Section 4) and continuous variable (Section 5) quantum states.

## 2. Classical regime of CPA

We consider a thin absorber as a symmetric 4-port device characterized by amplitude transmission $t$ and reflection $r$ coefficients with [4]

$$t = -r = 1/2. \tag{2.1}$$

The intensity absorption coefficient under single side illumination is $1 - |t|^2 - |r|^2 = 1/2$. Two counter-propagating along $z$-axis classical waves, Fig. 1(a), of the same polarization and frequency $\omega$,

$$E_k^{(in)} = \mathcal{E}_k^{(in)} e^{i(kz-\omega\tau)} + c.c., \tag{2.2}$$

$$E_{-k}^{(in)} = \mathcal{E}_{-k}^{(in)} e^{-i(kz+\omega\tau)} + c.c., \tag{2.3}$$

illuminate the absorber. Here $k = 2\pi/\lambda$ is a wave number, $\lambda$ is a wavelength, $\tau$ is a time variable and $\mathcal{E}_k^{(in)}$ and $\mathcal{E}_{-k}^{(in)}$ are complex amplitudes. Amplitudes $\mathcal{E}_k^{(out)}$ and $\mathcal{E}_{-k}^{(out)}$ of the output travelling waves, Fig. 1(d),

$$E_k^{(out)} = \mathcal{E}_k^{(out)} e^{i(kz-\omega\tau)} + c.c.,$$

$$E_{-k}^{(out)} = \mathcal{E}_{-k}^{(out)} e^{-i(kz+\omega\tau)} + c.c.,$$

are expressed through CPA-transformation,

$$\begin{pmatrix} \mathcal{E}_k^{(out)} \\ \mathcal{E}_{-k}^{(out)} \end{pmatrix} = H_{CPA}^{TR} \begin{pmatrix} \mathcal{E}_k^{(in)} \\ \mathcal{E}_{-k}^{(in)} \end{pmatrix},$$

where non-unitary matrix $H_{CPA}^{TR}$ describes the travelling waves scattering at the absorber (2.1),

$$H_{CPA}^{TR} = \frac{1}{2}\begin{pmatrix} 1 & -1 \\ -1 & 1 \end{pmatrix}.$$

Dissipation explicitly contains in the scattering matrix of the absorber and it is no restrictions on the energy conservation. For the input waves with amplitudes of equal moduli $\mathcal{E}_0$,

$$\mathcal{E}_k^{(in)} = \mathcal{E}_0 \cdot e^{i\theta_k}, \tag{2.4}$$

$$\mathcal{E}_{-k}^{(in)} = \mathcal{E}_0 \cdot e^{i\theta_{-k}}, \tag{2.5}$$

the regime of total absorption,

$$\mathcal{E}_k^{(out)} = \mathcal{E}_{-k}^{(out)} = 0, \tag{2.6}$$

holds if phase difference $\Delta\theta \equiv \theta_k - \theta_{-k} = 2\pi n$, $(n = 0,1,2,...)$, while the regime of total transmission,

$$\mathcal{E}_k^{(out)} = \mathcal{E}_k^{(in)}, \qquad \mathcal{E}_{-k}^{(out)} = \mathcal{E}_{-k}^{(in)}, \tag{2.7}$$

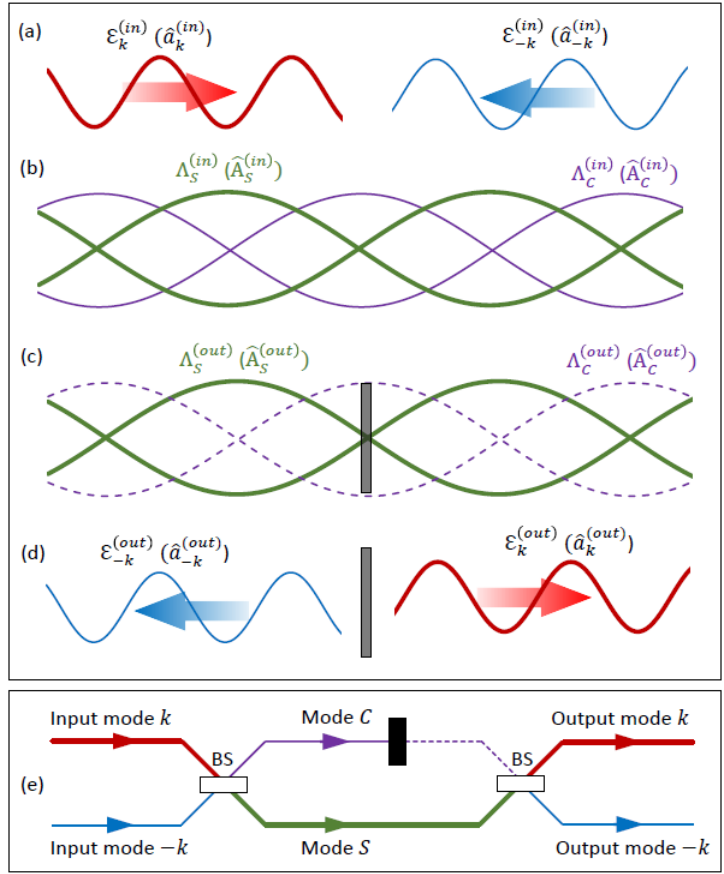

FIG. 1. **CPA in standing waves picture**. Input travelling waves (thick red and thin blue lines in (a)) are decomposed into superposition of two standing waves (thick green and thin purple lines in (b)). One of the standing waves (dashed purple line in (c)) is fully absorbed and the second one (solid green line in (c)) is not affected. The survived standing wave feeds up two output travelling waves (thin blue and thick red lines in (d)). This process is shown in the form of a diagram in (e): two input modes $k$ and $-k$ are transformed into modes $C$ and $S$ by the first BS-transformation; mode $C$ is fully absorbed by a beam block (black rectangular); the second BS-transformation results in output modes $k$ and $-k$. Colors and line styles of modes in (e) agree with corresponding colors and line styles of optical modes in (a)-(d). This picture is valid for both classical (amplitudes without hats) and quantum (operators in parenthesis) fields evolution.

holds if $\Delta\theta = (2n+1)\pi$. To explain these phenomena, we consider the total input field,

$$E^{(in)} = E_k^{(in)} + E_{-k}^{(in)}, \tag{2.8}$$

as a superposition of two – cosine and sine – standing waves, Fig. 1(b),

$$E^{(in)} = \sqrt{2}[\Lambda_C^{(in)} \cos kz + i\Lambda_S^{(in)} \sin kz]e^{-i\omega\tau} + c.c.. \tag{2.9}$$

Eq. (2.9) is derived by substituting (2.2) and (2.3) into (2.8) and using Euler's formula. It follows from (2.8) and (2.9), that the travelling and standing wave amplitudes are linked through the beamsplitter-like transformation (BS-transformation),

$$\begin{pmatrix} \Lambda_C^{(in)} \\ \Lambda_S^{(in)} \end{pmatrix} = H_{BS} \begin{pmatrix} \mathcal{E}_k^{(in)} \\ \mathcal{E}_{-k}^{(in)} \end{pmatrix}, \tag{2.10}$$

with the unitary matrix

$$H_{BS} = \frac{1}{\sqrt{2}}\begin{pmatrix} 1 & 1 \\ 1 & -1 \end{pmatrix}. \tag{2.11}$$

From (2.4), (2.5), (2.10) and (2.11), the standing wave amplitudes are expressed as

$$\Lambda_C^{(in)} = \sqrt{2}\mathcal{E}_0 e^{i(\theta_k+\theta_{-k})/2} \cos\frac{\Delta\theta}{2},$$

$$\Lambda_S^{(in)} = i\sqrt{2}\mathcal{E}_0 e^{i(\theta_k+\theta_{-k})/2} \sin\frac{\Delta\theta}{2},$$

and one may redistribute the energy between the standing waves by changing the relative phase of the input waves $\Delta\theta$. Since the cosine wave is formed by the in-phase combination of the input travelling waves,

$$\cos kz \sim e^{ikz} + e^{-ikz}, \tag{2.12}$$

it is fully absorbed (compare with (2.4)). The sine standing wave, in contrast, is formed by the out-of-phase component of the input travelling waves,

$$\sin kz \sim e^{ikz} - e^{-ikz}, \tag{2.13}$$

and it is not affected by the absorber (compare with (2.7)). Similar to (2.8) and (2.9), the total output field can be written as

$$E^{(out)} = E_k^{(out)} + E_{-k}^{(out)} = \sqrt{2}\left[\Lambda_C^{(out)} \cos kz + i\Lambda_S^{(out)} \sin kz\right] e^{-i\omega\tau} + c.c.,$$

where the standing-to-travelling waves transformation is expressed as

$$\begin{pmatrix} \mathcal{E}_k^{(out)} \\ \mathcal{E}_{-k}^{(out)} \end{pmatrix} = H_{BS} \begin{pmatrix} \Lambda_C^{(out)} \\ \Lambda_S^{(out)} \end{pmatrix}.$$

According to (2.12) and (2.13), CPA transformation of the standing waves, Fig. 1(c), is read as

$$\begin{pmatrix} \Lambda_C^{(out)} \\ \Lambda_S^{(out)} \end{pmatrix} = H_{CPA}^{ST} \begin{pmatrix} \Lambda_C^{(in)} \\ \Lambda_S^{(in)} \end{pmatrix}$$

with the non-unitary matrix

$$H_{CPA}^{ST} = \begin{pmatrix} 0 & 0 \\ 0 & 1 \end{pmatrix}. \tag{2.14}$$

Finally, the classical regime of CPA is described as a result of the following consecutive transformations:
- two counter-propagating travelling waves, Fig. 1(a), form two standing waves, Fig. 1(b);
- the cosine standing wave is fully absorbed, while the sine wave is not affected, Fig. 1(c);
- the survived sine wave is split into two output travelling waves, Fig. 1(d).

This sequence of transformations can be written in a matrix form as

$$\begin{pmatrix} \mathcal{E}_k^{(out)} \\ \mathcal{E}_{-k}^{(out)} \end{pmatrix} = H_{BS} \cdot H_{CPA}^{ST} \cdot H_{BS} \begin{pmatrix} \mathcal{E}_k^{(in)} \\ \mathcal{E}_{-k}^{(in)} \end{pmatrix}, \tag{2.15}$$

where $H_{BS} \cdot H_{CPA}^{ST} \cdot H_{BS} = H_{CPA}^{TR}$. It is convenient to depict this process in a diagram form, Fig. 1(e). Here two input modes, $k$ (upper red line) and $-k$ (bottom blue line), are mixed on the first 'virtual' beamsplitter (BS). The upper output mode $C$ of the BS (purple line) is fully absorbed, while the bottom mode $S$ (green line) feeds up one of the input ports of the second 'virtual' BS resulting in output modes $k$ (upper red line) and $-k$ (bottom blue line).

The standing wave picture leads us to important conclusion that the absorber (2.1) can be implemented in a subwavelength design only. We refer to the fact that the absorber can operate in the regime of total transmission (2.7). Obviously, it can be achieved iff the absorber is placed at node of the standing wave which requires a subwavelength thickness. This conclusion is also valid for the absorber with $t = r = 1/2$ [15,16,24] which has the opposite functionality: it is opaque for the sine standing wave and transparent for the cosine standing wave.

## 3. Quantum regime of CPA

Theory of quantum regime of CPA can be built similar to classical description. Electric field of the input and output travelling light modes are expressed now through the corresponding operators [25,26]:

$$\hat{E}_k^{(in)} = i\mathcal{E}^{(q)} \cdot \hat{a}_k^{(in)} e^{i(kz-\omega\tau)} + H.c.,$$

$$\hat{E}_{-k}^{(in)} = i\mathcal{E}^{(q)} \cdot \hat{a}_{-k}^{(in)} e^{-i(kz+\omega\tau)} + H.c.,$$

$$\hat{E}_k^{(out)} = i\mathcal{E}^{(q)} \cdot \hat{a}_k^{(out)} e^{i(kz-\omega\tau)} + H.c.,$$

$$\hat{E}_{-k}^{(out)} = i\mathcal{E}^{(q)} \cdot \hat{a}_{-k}^{(out)} e^{-i(kz+\omega\tau)} + H.c.,$$

where $\mathcal{E}^{(q)} = \sqrt{\hbar\omega/2\varepsilon_0 V}$, $\hbar$ is a reduced Planck constant, $\varepsilon_0$ is a vacuum permittivity and $V$ is a quantization volume, and mode amplitudes satisfy standard commutation relations:

$$[\hat{a}_k^{(in)}, \hat{a}_k^{(in)\dagger}] = [\hat{a}_{-k}^{(in)}, \hat{a}_{-k}^{(in)\dagger}] = [\hat{a}_k^{(out)}, \hat{a}_k^{(out)\dagger}] = [\hat{a}_{-k}^{(out)}, \hat{a}_{-k}^{(out)\dagger}] = 1,$$

$$[\hat{a}_k^{(in)}, \hat{a}_{-k}^{(in)\dagger}] = [\hat{a}_k^{(out)}, \hat{a}_{-k}^{(out)\dagger}] = 0, \qquad (3.1)$$

and so on. Transformation of the travelling waves on the absorber is non-unitary and, according to the fluctuation-dissipation theorem, noise operators, $\hat{f}_k$ and $\hat{f}_{-k}$, are appended in order to validate (3.1),

$$\begin{pmatrix} \hat{a}_k^{(out)} \\ \hat{a}_{-k}^{(out)} \end{pmatrix} = H_{CPA}^{TR} \begin{pmatrix} \hat{a}_k^{(in)} \\ \hat{a}_{-k}^{(in)} \end{pmatrix} + \begin{pmatrix} \hat{f}_k \\ \hat{f}_{-k} \end{pmatrix},$$

where commutation relations of noise operators, $[\hat{f}_k, \hat{f}_k^\dagger] = [\hat{f}_{-k}, \hat{f}_{-k}^\dagger] = 1 - |t|^2 - |r|^2$, are absorption dependent. Next, we express total input $\hat{E}^{(in)}$ and output $\hat{E}^{(out)}$ fields in the basis of quantized standing waves:

$$\hat{E}^{(in)} = \hat{E}_k^{(in)} + \hat{E}_{-k}^{(in)} = \mathcal{E}_{ST}^{(q)} [\hat{A}_C^{(in)} \cos kz + i\hat{A}_S^{(in)} \sin kz] e^{-i\omega\tau} + H.c.,$$

$$\hat{E}^{(out)} = \hat{E}_k^{(out)} + \hat{E}_{-k}^{(out)} = \mathcal{E}_{ST}^{(q)} [\hat{A}_C^{(out)} \cos kz + i\hat{A}_S^{(out)} \sin kz] e^{-i\omega\tau} + H.c.,$$

where $\mathcal{E}_{st}^{(q)} = i\sqrt{2}\mathcal{E}^{(q)}$ and

$$\begin{pmatrix} \hat{A}_C^{(in)} \\ \hat{A}_S^{(in)} \end{pmatrix} = H_{BS} \begin{pmatrix} \hat{a}_k^{(in)} \\ \hat{a}_{-k}^{(in)} \end{pmatrix}, \qquad (3.2)$$

$$\begin{pmatrix} \hat{a}_k^{(out)} \\ \hat{a}_{-k}^{(out)} \end{pmatrix} = H_{BS} \begin{pmatrix} \hat{A}_C^{(out)} \\ \hat{A}_S^{(out)} \end{pmatrix} \qquad (3.3)$$

with the matrix (2.11). Standard commutation relations stay valid for amplitudes of the standing waves:

$$[\hat{A}_C^{(in)}, \hat{A}_C^{(in)\dagger}] = [\hat{A}_S^{(in)}, \hat{A}_S^{(in)\dagger}] = [\hat{A}_C^{(out)}, \hat{A}_C^{(out)\dagger}] = [\hat{A}_S^{(out)}, \hat{A}_S^{(out)\dagger}] = 1,$$

$$[\hat{A}_C^{(in)}, \hat{A}_S^{(in)\dagger}] = [\hat{A}_C^{(out)}, \hat{A}_S^{(out)\dagger}] = 0$$

and so on. By using (3.2) and (3.3), we may trace evolution of the quantized standing waves:

$$\widehat{A}_C^{(out)} = \frac{\hat{f}_k + \hat{f}_{-k}}{\sqrt{2}} \equiv \hat{F}_C, \qquad (3.4)$$

$$\widehat{A}_S^{(out)} = \frac{\hat{a}_k^{(in)} - \hat{a}_{-k}^{(in)}}{\sqrt{2}} = \widehat{A}_S^{(in)}, \qquad (3.5)$$

where only the noise operators contribute to the output amplitude of the cosine wave (it ends up in a vacuum state), while the sine wave experiences no changes. Full absorption of the cosine wave imposes standard bosonic commutation relation on the noise operator $\hat{F}_C$: $[\hat{F}_C, \hat{F}_C^\dagger] = 1$. Boson-like behaviour of the noise operator enables full exchange of the quantum states between the cosine wave and the absorber allowing quantum light manipulation as it will be discussed in the following sections. Transformations (3.4) and (3.5) can be written in a matrix form with the matrix (2.14),

$$\begin{pmatrix} \widehat{A}_C^{(out)} \\ \widehat{A}_S^{(out)} \end{pmatrix} = H_{CPA}^{ST} \begin{pmatrix} \widehat{A}_C^{(in)} \\ \widehat{A}_S^{(in)} \end{pmatrix} + \begin{pmatrix} \hat{F}_C \\ 0 \end{pmatrix}. \qquad (3.6)$$

Thus, to describe quantum regime of CPA we use the following procedure:
- switch the travelling waves basis, Fig. 1(a), to the standing waves basis, Fig. 1(b);
- the cosine standing wave is fully absorbed while the sine wave is not affected, Fig. 1(c);
- mixture of the cosine wave in a vacuum state and the sine wave in its initial state forms two output travelling waves, Fig. 1(d).

Due to the presence of the noise operator column in (3.6), CPA transformation of quantum light cannot be written in the form of matrix multiplication as (2.15) for classical light. Nevertheless, sequence of transformations is the same: beamsplitter-like transformation (3.2) followed by CPA-transformation (3.6) and, finally, by beamsplitter-like transformation (3.3). Accordingly, the diagram in Fig. 1(e) is a valid representation of the quantum regime of CPA, where the upper input port of the second beamsplitter is fed up with a vacuum state now.

Developed here theory of coherent perfect absorption is based on representation of monochromatic waves as a combination of standing waves. In practise, the absorber is illuminated by the wave packets rather than monochromatic waves. Generalization of the theory for light modes of finite duration can be done in the following way. A pair of continuum amplitude operators $\hat{a}_{k'}^{(in)}(\omega')$ and $\hat{a}_{-k'}^{(in)}(\omega')$ should be defined for each spectral component $\omega'$ with the wave numbers of opposite sign, $|-k'| = k' = c/\omega'$. Each pair of waves with amplitudes $\hat{a}_{k'}^{(in)}(\omega')$ and $\hat{a}_{-k'}^{(in)}(\omega')$ excites the corresponding pair of standing waves where the cosine wave is dissipated while the sine wave is not affected. Dispersion (dependence of optical response on frequency) of the absorber should be considered [15]. If dispersion is negligible, single mode theory developed here holds. Dispersion of the typical coherent perfect absorbers used in experiments is indeed negligible and starts playing noticeable role for very short (≲10-20 fs) pulses [27].

Deviation of the absorber optical response from (2.1) does not change transformation of the sine wave amplitude, $\widehat{A}_S^{(out)} \sim (t - r)\widehat{A}_S^{(in)}$, since for any subwavelength absorber relation $t = 1 + r$ is valid [10,28] (absence of interaction at the standing wave node). In contrast, transformation of the cosine wave amplitude, $\widehat{A}_C^{(out)} \sim (t + r)\widehat{A}_C^{(in)}$, changes, and its output state will be a mixture of input and vacuum states with corresponding weights.

As it will be shown in the following sections, quantum regimes of CPA may lead to generation of entanglement between light and absorber. To observe and exploit these phenomena, a system (atoms, plasmons etc.) able to store quantum states, rather than just dissipate the energy, should be used as the absorber [21].

## 4. CPA of discrete variable quantum states

Depending on variables used, quantum optics and quantum information appears in two forms: discrete and continuous [29,30]. Discrete variable (DV) consideration is valid if operators of interest have discrete spectrum of eigenvalues – energy of quantized oscillator and polarization of light are vivid examples. Dense coding [31,32], quantum teleportation [33,34], quantum key distribution [35,36], quantum information processing [37-39], quantum memory [40-45] and quantum metrology [46] were proposed and demonstrated by exploiting DV states, as well as an ability to realize quantum information protocols in integrated design [47-49]. CPA of DV quantum states may provide novel approaches for quantum light control [10,11,22], manipulation of quantum states of light [12,13,18] and light-matter interaction [21]. In this section, we apply the developed theory of CPA to the case when DV quantum states of light are used.

### *4.1 CPA of single photon states*

CPA of a single photon where input travelling waves carrying the path-entangled state

$$|\Psi\rangle_{k,-k} = \frac{1}{\sqrt{2}}\left(|1\rangle_k|0\rangle_{-k} + e^{i\Delta\theta}|0\rangle_k|1\rangle_{-k}\right). \tag{4.1}$$

Here term $|1\rangle_k|0\rangle_{-k}$ describes a single photon presented in mode $k$ and a vacuum state in mode $-k$, and so on, and $\Delta\theta$ is a phase difference between the wave function components. Since the standing waves appear as a result of the BS-transformation (3.2) of the travelling waves, we use the following decomposition:

$$|1\rangle_k|0\rangle_{-k} \rightarrow \frac{1}{\sqrt{2}}(|1\rangle_C|0\rangle_S + |0\rangle_C|1\rangle_S), \tag{4.2}$$

$$|0\rangle_k|1\rangle_{-k} \rightarrow \frac{1}{\sqrt{2}}(|1\rangle_C|0\rangle_S - |0\rangle_C 1\rangle_S). \tag{4.3}$$

By substituting them into (4.1), we find the wave function of the standing waves (with accuracy to a common phase):

$$|\Phi\rangle_{C,S} = \cos\frac{\Delta\theta}{2}|1\rangle_C|0\rangle_S - i\sin\frac{\Delta\theta}{2}|0\rangle_C|1\rangle_S.$$

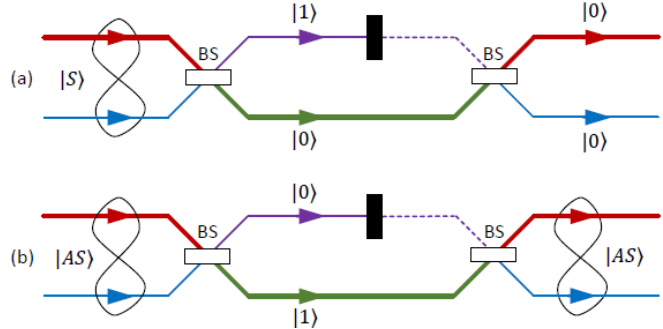

FIG. 2. **CPA of single photon states**. (a) Symmetric state of the input modes $k$ and $-k$ (input travelling waves) is coupled to the mode $C$ (the cosine standing wave) which is fully absorbed. (b) Anti-symmetric state of the input modes $k$ and $-k$ is coupled to the mode $S$ (the sine standing wave) which escapes dissipation and is coupled back to anti-symmetric state of the output modes $k$ and $-k$.

When the input travelling waves are in a symmetric superposition state ($\Delta\theta = 2\pi n$), the photon is coupled to the cosine standing wave only. The regime of deterministic photon absorption takes place, Fig. 2(a). Opposite, an anti-symmetric superposition state ($\Delta\theta = (2n+1)\pi$) of the travelling waves excites the sine standing wave only, which experiences no losses and it is transformed back into the output modes $k$ and $-k$, Fig. 2(b),

$$|0\rangle_C|1\rangle_S \to \frac{1}{\sqrt{2}}(|1\rangle_k|0\rangle_{-k} - |0\rangle_k|1\rangle_{-k}). \tag{4.4}$$

In general case, photon is coherently distributed between the cosine and sine waves with the probability of $\cos^2(\Delta\theta/2)$ to be absorbed. Since the only way to pass through the absorber is by coupling to the sine standing wave, the output photon can leave the absorber in the state (4.4) only. In this sense, coherent perfect absorber works as a quantum state filter, transmitting only anti-symmetric part of the wavefunction. Absorber with $t = r = 1/2$ transmits symmetric and absorbs anti-symmetric components of the wave function.

*4.2 CPA of two photon states*

Two different second quantization formalisms – either bosonic [15] or fermionic [13] – have been used to describe CPA of two-photon states. Here, we consider CPA of two photon states under a single theory of CPA of quantized standing waves.

All possible states of two photons, photon *A* and photon *B*, occupying two travelling wave modes (mode *k* and mode *-k*) can be expressed through Bell states [50]:

$$|\psi^{(+)}\rangle_{k,-k} = \frac{1}{\sqrt{2}}(|A\rangle_k|B\rangle_{-k} + |B\rangle_k|A\rangle_{-k}), \tag{4.5}$$

$$|\psi^{(-)}\rangle_{k,-k} = \frac{1}{\sqrt{2}}(|A\rangle_k|B\rangle_{-k} - |B\rangle_k|A\rangle_{-k}), \tag{4.6}$$

$$|\varphi^{(+)}\rangle_{k,-k} = \frac{1}{\sqrt{2}}(|A\rangle_k|B\rangle_k + |A\rangle_{-k}|B\rangle_{-k}), \tag{4.7}$$

$$|\varphi^{(-)}\rangle_{k,-k} = \frac{1}{\sqrt{2}}(|A\rangle_k|B\rangle_k - |A\rangle_{-k}|B\rangle_{-k}). \tag{4.8}$$

Here combination $|A\rangle_k|B\rangle_{-k}$ reads as 'photon *A* is presented in mode *k* and photon *B* is presented in mode *-k*', and so on. States (4.5) and (4.6) describe scenarii when one photon is coming from each side of the absorber, and their form reflects symmetrisation postulate of quantum mechanics. States (4.7) and (4.8) are *NOON*-states with *N=2*. States $|\psi^{(+)}\rangle$, $|\varphi^{(+)}\rangle$ and $|\varphi^{(-)}\rangle$ possess a bosonic symmetry, while state $|\psi^{(-)}\rangle$ has a fermionic symmetry and should be accompanied by an anti-symmetric polarization (or other degree of freedom) part of the wave function so that the latter has a bosonic symmetry, as it should be for the photons. BS-transformation of the states (4.5)-(4.8) can be found similar to (4.2) and (4.3), where

$$|A\rangle_{\pm k} \to \frac{1}{\sqrt{2}}(|A\rangle_C \pm |A\rangle_S), \tag{4.9}$$

$$|B\rangle_{\pm k} \to \frac{1}{\sqrt{2}}(|B\rangle_C \pm |B\rangle_S). \tag{4.10}$$

By substituting (4.9) and (4.10) into (4.5)-(4.8), we find the corresponding states of the standing waves:

$$|\psi^{(+)}\rangle_{k,-k} \to |\varphi^{(-)}\rangle_{C,S},$$

$$|\psi^{(-)}\rangle_{k,-k} \to |\psi^{(-)}\rangle_{C,S},$$

$$|\varphi^{(+)}\rangle_{k,-k} \to |\varphi^{(+)}\rangle_{C,S},$$

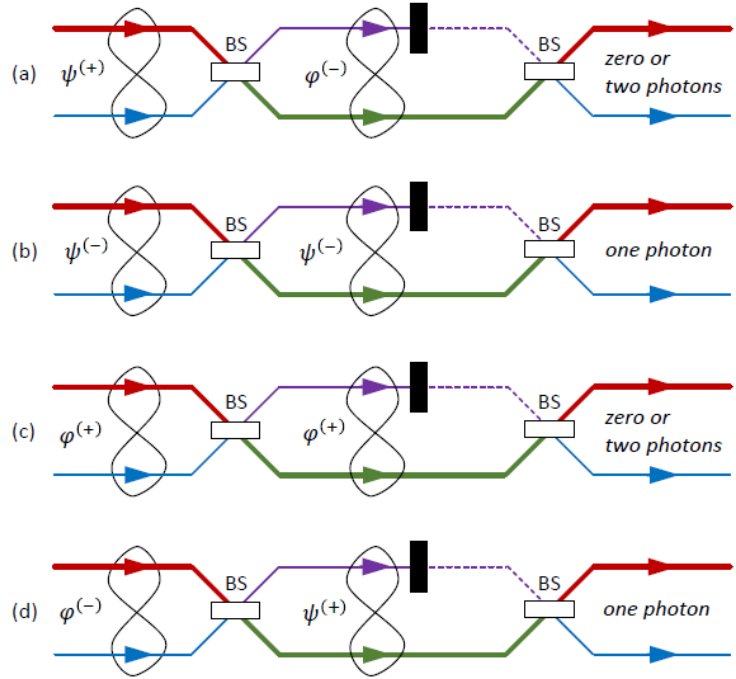

FIG. 3. **CPA of two photon Bell states**. The result of CPA is explained by Bell states transformation. If the output of the first BS-transformation is the *NOON* state, probabilistic two photon absorption follows, (a) and (c). In contrast, if one photon is always presented in output ports of the first BS, deterministic one photon absorption takes place, (b) and (d).

$$\left|\varphi^{(-)}\right\rangle_{k,-k} \to \left|\psi^{(+)}\right\rangle_{C,S},$$

where, for instance, $\left|\varphi^{(-)}\right\rangle_{C,S} = 1/\sqrt{2}(|A\rangle_C|B\rangle_C - |A\rangle_S|B\rangle_S)$ and so on.

Similar to boson bunching in Hong-Ou-Mandel effect [51], the two-photon state $\left|\psi^{(+)}\right\rangle$ of the travelling waves is transformed into the *NOON*(-) state $\left|\varphi^{(-)}\right\rangle$ of the standing waves, Fig. 3(a). Since the cosine wave is fully absorbed, probabilistic zero or two photon absorption follows. If two photons survive (both photon are 'passing through' the sine wave), the output modes state can be found straightforward:

$$|A\rangle_S|B\rangle_S \to \frac{1}{2}(|A\rangle_k|B\rangle_k + |A\rangle_{-k}|B\rangle_{-k} - |A\rangle_k|B\rangle_{-k} - |A\rangle_{-k}|B\rangle_k), \qquad (4.11)$$

where with 50% probability both photons are found in the same mode and with 50% probability – in different modes.

The fermionic state $\left|\psi^{(-)}\right\rangle$ of the travelling waves is an eigenstate of any BS-transformation (in agreement with Pauli exclusion principle), and it excites the fermionic state of the standing waves, Fig. 3(b). One photon is always presented in the cosine wave and it is absorbed. The second photon is coupled to the sine wave which, in turn, excites the single photon anti-symmetric state of the output modes, similar to (4.4). The same conclusion is valid for the input state $\left|\varphi^{(-)}\right\rangle$ which is transformed into $\left|\psi^{(+)}\right\rangle$ state of the standing waves, Fig. 3(d). Again, one photon is always presented in the cosine wave and the second one is presented in the sine wave.

The *NOON*(+) state $\left|\varphi^{(+)}\right\rangle$ is an eigenstate of the matrix (2.11) and it is preserved during the BS-transformation, Fig. 3(c). The *NOON*(+) state of the standing waves exhibit similar to the *NOON*(-) state behaviour with 50% probability of two photon absorption and 50% probability to find both photons in the output modes $k$ and $-k$ in the state (4.11).

For all two-photon Bell-state we get 50% of the average light absorption. At the same time, statistics of the individual events differs from probabilistic zero- or two-photon absorption to deterministic one photon absorption depending on the input state. These phenomena were demonstrated experimentally in Refs. [12,14,20].

*4.3 CPA of NOON states with arbitrary N*

In this section, we introduce the regime of CPA of *NOON* state

$$|NOON\rangle_{k,-k} = \frac{1}{\sqrt{2}}\left(|N\rangle_k|0\rangle_{-k} + e^{i\Delta\theta}|0\rangle_k|N\rangle_{-k}\right),$$

with arbitrary number of entangled photons *N*. Here the term $|N\rangle_k|0\rangle_{-k}$ stands for *N* photons in the mode *k*, and zero photons in the mode *-k*, and so on, and $\Delta\theta$ is a phase difference between two terms. By rewriting the *NOON* state as

$$|NOON\rangle_{k,-k} = \frac{1}{\sqrt{2}\sqrt{N!}}\left(\left\{\hat{a}_k^{(in)\dagger}\right\}^N + e^{i\Delta\theta}\left\{\hat{a}_{-k}^{(in)\dagger}\right\}^N\right)|0\rangle_k|0\rangle_{-k},$$

and using the reverse BS-transformation (3.2), we get the input state of the standing waves [52]:

$$|NOON\rangle_{k,-k} \rightarrow \frac{1}{\sqrt{2^{N-1}}}\sum_{m=0}^{N}\binom{N}{m}^{1/2}\cos\left\{\frac{1}{2}(\pi m + \Delta\theta)\right\}|N-m\rangle_c|m\rangle_s.$$

Here $\binom{N}{m} = \frac{N!}{(N-m)!m!}$ is a binomial coefficient and the term $|N-m\rangle_c|m\rangle_s$ describes the state with $N-m$ photons in the cosine standing wave and $m$ photons in the sine standing wave. This representation allows us to analyse the outcome of CPA process. For instance, the three photon *NOON* state of the counter-propagating waves is transformed into

$$\frac{1}{2}\left\{\cos\frac{\Delta\theta}{2}\left(|3\rangle_c|0\rangle_s - \sqrt{3}|1\rangle_c|2\rangle_s\right) + \sin\frac{\Delta\theta}{2}\left(|0\rangle_c|3\rangle_s - \sqrt{3}|2\rangle_c|1\rangle_s\right)\right\}$$

state of the standing waves. Thus, if $\Delta\theta = 2\pi n$, three photon absorption takes place with the probability of 25%, while one photon is absorbed with the probability of 75%. If $\Delta\theta = (2n+1)\pi$, either zero or two photons are absorbed, where two photon absorption takes place with the probability of 75% – higher than for *NOON(+)* state with *N=2* [14]. For the four photon *NOON* state of the counter-propagating input waves, the standing waves state is

$$\frac{1}{2\sqrt{2}}\left\{\cos\frac{\Delta\theta}{2}\left(|4\rangle_c|0\rangle_s + |0\rangle_c|4\rangle_s - \sqrt{6}|2\rangle_c|2\rangle_s\right) + 2\sin\frac{\Delta\theta}{2}\left(|3\rangle_c|1\rangle_s - |1\rangle_c|3\rangle_s\right)\right\}.$$

Probabilistic two or four photon absorption takes place if $\Delta\theta = 2\pi n$ (with probabilities of 12.5% and 75%, accordingly) and probabilistic one or three photon absorption takes place if $\Delta\theta = (2n+1)\pi$ (with equal 50% probabilities). It is straightforward to get the state of the output light modes *k* and *-k* by applying the second BS-transformation to the state of the sine standing wave.

In a similar manner CPA of *NOON* states with higher *N* and other DV quantum states, such as multiphoton Fock [53], Greenberger–Horne–Zeilinger [54], mixed etc. states can be described, which we leave out of the scope of this paper.

## 5. CPA of continuous variable quantum states

Parallel branch of quantum information deals with continuous variables (CV) states [29,30] where operators of interest have continuous spectra of eigenvalues. Quadrature amplitudes of quantum light and collective spin of atomic ensemble are among widely used examples [55]. Protocols of quantum teleportation [56-59], quantum information processing [60,61], quantum communication [62] and quantum memory [63-69] were proposed and demonstrated with CV states, as well as an ability to implement them in integrated platforms [70-72]. Similar to CPA of DV quantum states, CPA of CV states may be used to control and manipulate quantum states of light [17,19] and to enable advanced protocols of light-matter interaction [21]. In this section, we analyse CPA of the most exploitable states in CV quantum information – squeezed and entangled states, and shortly outline CPA of other CV states. Since CPA of CV quantum states has not been analysed from the states evolution point of view, we, first, remind major properties of CV quantum states and their representation in a phase space, and it can be found in more details somewhere else [26,29,30,73,74].

*5.1 CV quantum states*

Each optical mode with amplitude $\hat{a} = (\hat{X}_1 + i\hat{X}_2)/2$ is associated with a quantized oscillator, where optical quadrature amplitudes, $\hat{X}_1$ and $\hat{X}_2$, have a meaning of oscillator position and momentum with commutation relation $[\hat{X}_1, \hat{X}_2] = 2i$. Variance of position and momentum, $(\Delta X_1)^2$ and $(\Delta X_2)^2$,

$$(\Delta X_i)^2 = \langle (\hat{X}_i)^2 \rangle - (\langle \hat{X}_i \rangle)^2, \quad i = 1,2,$$

are linked through Heisenberg uncertainty relation, $(\Delta X_1)^2 (\Delta X_2)^2 \geq 1$, and can be measured by homodyne detection [26,75]. CV quantum state is reflected in a phase space by a 'coherent' arrow with coordinate $(\langle \hat{X}_1 \rangle, \langle \hat{X}_2 \rangle)$ accompanied by the uncertainty region. For instance, a coherent state $|\alpha\rangle$ (where $\alpha = |\alpha|e^{i\theta}$ is a complex number) is represented by a 'coherent' arrow with coordinate $(2|\alpha|\cos\theta, 2|\alpha|\sin\theta)$ and uncertainty, or noise region shaped into circle with radius 1 ($\Delta X_1 \equiv \sqrt{(\Delta X_1)^2} = 1, \Delta X_2 \equiv \sqrt{(\Delta X_2)^2} = 1$). The sum of 'coherent' arrow with a random arrow within uncertainty region defines particular realization of the state in a single trial [76]. Uncertainty region of the squeezed states is represented by the ellipse with one of the radii smaller than 1 where for ideal squeezing $(\Delta X_1)^2 (\Delta X_2)^2 = 1$ and for non-ideal squeezing $(\Delta X_1)^2 (\Delta X_2)^2 > 1$ [26].

We express amplitudes of the input counter-propagating and standing waves through their quadrature components:

$$\hat{a}_k^{(in)} = \frac{\hat{X}_{k,1}^{(in)} + i\hat{X}_{k,2}^{(in)}}{2}, \quad \hat{a}_{-k}^{(in)} = \frac{\hat{X}_{-k,1}^{(in)} + i\hat{X}_{-k,2}^{(in)}}{2},$$

$$\hat{A}_C^{(in)} = \frac{\hat{X}_{C,1}^{(in)} + i\hat{X}_{C,2}^{(in)}}{2}, \quad \hat{A}_S^{(in)} = \frac{\hat{X}_{S,1}^{(in)} + i\hat{X}_{S,2}^{(in)}}{2},$$

where the BS-transformation (3.2) is expressed now as

$$\hat{X}_{C,i}^{(in)} = \frac{\hat{X}_{k,i}^{(in)} + \hat{X}_{-k,i}^{(in)}}{\sqrt{2}}, \quad \hat{X}_{S,i}^{(in)} = \frac{\hat{X}_{k,i}^{(in)} - \hat{X}_{-k,i}^{(in)}}{\sqrt{2}},$$

where $i = 1,2$. As it was discussed in Section 3, the sine standing wave is not affected by the absorber, and

$$\hat{X}_{S,i}^{(out)} = \hat{X}_{S,i}^{(in)},$$

while the cosine wave amplitude is replaced by the noise operator (3.4), and

$$\hat{X}_{C,i}^{(out)} = \hat{X}_{F,i}.$$

Here $\hat{F}_C = (\hat{X}_{F,1} + i\hat{X}_{F,2})/2$ and a vacuum state is associated with this amplitude. Finally, the output travelling wave amplitudes are expressed as

$$\hat{a}_k^{(out)} = \frac{\hat{X}_{k,1}^{(out)} + i\hat{X}_{k,2}^{(out)}}{2}, \quad \hat{a}_{-k}^{(out)} = \frac{\hat{X}_{-k,1}^{(out)} + i\hat{X}_{-k,2}^{(out)}}{2},$$

and, according to the second BS-transformation (3.3),

$$\hat{X}_{k,i}^{(out)} = \frac{\hat{X}_{F,i} + \hat{X}_{S,i}^{(out)}}{\sqrt{2}}, \quad \hat{X}_{-k,i}^{(out)} = \frac{\hat{X}_{F,i} - \hat{X}_{S,i}^{(out)}}{\sqrt{2}}. \tag{5.1}$$

To quantify the amount of entanglement between any two modes, operators of relative position and total momentum are introduced (operators of total position and relative momentum can be used instead). For instance, relative position $\hat{Q}_{k,-k}$ and total momentum $\hat{P}_{k,-k}$ of the input counter-propagating waves, are defined as

$$\hat{Q}_{k,-k} = \frac{\hat{X}_{k,1}^{(in)} - \hat{X}_{-k,1}^{(in)}}{\sqrt{2}}, \tag{5.2}$$

$$\hat{P}_{k,-k} = \frac{\hat{X}_{k,2}^{(in)} + \hat{X}_{-k,2}^{(in)}}{\sqrt{2}}. \tag{5.3}$$

As a measure of entanglement between two modes, we use inseparability parameter $S_{k,-k}$ [77]:

$$S_{k,-k} = \left(\Delta Q_{k,-k}\right)^2 + \left(\Delta P_{k,-k}\right)^2.$$

If two optical modes are in a coherent state, they are characterized by the quantum shot noise with $S_{k,-k} = 2$. For separable states, inseparability parameter is equal or exceeds the shot noise,

$$S_{k,-k} \geq 2,$$

while for inseparable, or entangled states strong correlations 'dump' the noise,

$$S_{k,-k} < 2.$$

In this section, we will distinguish between intensity and coherence absorption [17]. Intensity of the absorbed light $I_{abs}$ for any input state is defined by intensity of the cosine wave,

$$I_{abs} = \langle \hat{A}_C^{(in)\dagger} \hat{A}_C^{(in)} \rangle = \frac{I_{in}}{2} + Re\left\{\langle \hat{a}_k^{(in)\dagger} \hat{a}_{-k}^{(in)} \rangle\right\}.$$

Here transformation (3.2) was used, and $I_{in} = I_k^{(in)} + I_{-k}^{(in)}$ is intensity of the input light ($I_k^{(in)} = \langle \hat{a}_k^{(in)\dagger} \hat{a}_k^{(in)} \rangle$ and $I_{-k}^{(in)} = \langle \hat{a}_{-k}^{(in)\dagger} \hat{a}_{-k}^{(in)} \rangle$ are intensities of each input mode). Thus, intensity absorption coefficient,

$$\mathcal{A}^I = \frac{I_{abs}}{I_{in}} = \frac{1}{2} + \frac{Re\left\{\langle \hat{a}_k^{(in)\dagger} \hat{a}_{-k}^{(in)} \rangle\right\}}{I_{in}}, \tag{5.4}$$

is defined by the correlations between the input modes, $\langle \hat{a}_k^{(in)\dagger} \hat{a}_{-k}^{(in)} \rangle$. It can be simplified for separable, non-entangled input states as

$$\mathcal{A}_{sep}^I = \frac{1}{2} + \frac{Re\left\{\langle \hat{a}_k^{(in)} \rangle^* \langle \hat{a}_{-k}^{(in)} \rangle\right\}}{I_{in}}, \tag{5.5}$$

where averaging of amplitude operators is done separately. Absorption coefficient of coherence (in Glauber's sense) is defined as

$$\mathcal{A}^C = \frac{C_{abs}}{C_{in}} = \frac{1}{2} + \frac{Re\left\{\langle \hat{a}_k^{(in)} \rangle^* \langle \hat{a}_{-k}^{(in)} \rangle\right\}}{C_{in}}. \tag{5.6}$$

Here $C_{in} = C_k^{(in)} + C_{-k}^{(in)}$ is the total input coherence with the corresponding coherence of each mode $C_k^{(in)} = |\langle \hat{a}_k^{(in)} \rangle|^2$ and $C_{-k}^{(in)} = |\langle \hat{a}_{-k}^{(in)} \rangle|^2$, and the absorbed coherence $C_{abs} = C_C = |\langle \hat{A}_C^{(in)} \rangle|^2$ is defined by the coherence of the cosine standing wave. Opposite to intensity absorption, $\mathcal{A}^C$ does not depend on correlations between the input modes. Coherence can be thought of as a measure of 'classical interference ability' of light: if two light modes are mixed on a beamsplitter the 'coherent' components will interfere in a classical sense as a phase-dependent redistribution of *average* light intensity. In contrast, the noise component of light, which does not contribute to coherence, will be split equally between the output ports of the beamsplitter. Coherence defines a length of the corresponding 'coherent' arrow in a phase space, where result of mixture of two arrows depends on their length and mutual orientation. We note, that intensity absorption coefficient in its general form (5.4) depends on both 'coherent' and noise components of light.

*5.2 CPA of squeezed states*

In this section, the input counter-propagating waves are assumed to be in coherent squeezed states, $|\alpha_k, \zeta_k\rangle_k$ and $|\alpha_{-k}, \zeta_{-k}\rangle_{-k}$, where $\alpha_k = |\alpha_k| e^{i\theta_k}$, $\alpha_{-k} = |\alpha_{-k}| e^{i\theta_{-k}}$, $\zeta_k = \xi_k e^{i\phi_k}$ and $\zeta_{-k} = \xi_{-k} e^{i\phi_{-k}}$ are complex numbers, and $\xi_k$ and $\xi_{-k}$ are squeezing parameters (real numbers). To depict the state $|\alpha_k, \zeta_k\rangle_k$ in a phase space, one may start with a 'coherent' arrow $(2|\alpha_k|\cos\theta_k, 2|\alpha_k|\sin\theta_k)$ and a unit radius uncertainty circle which are then squeezed in rotated by $\phi_k/2$ phase space coordinate frame [26,73]. To find coefficients of intensity and coherence absorption, (5.5) and (5.6) should be solved with the input (separable) states $|\alpha_k, \zeta_k\rangle_k$ and $|\alpha_{-k}, \zeta_{-k}\rangle_{-k}$. Detailed analysis of these coefficients as a function of the input state parameters was done in [17], and we do not replicate the calculations here. Instead, we consider the problem from the state evolution point of view which allows us to explain the results of squeezed light interference on coherent perfect absorber. We start our consideration with an important case of two identical input squeezed states:

$$\alpha_k = \alpha_{-k} = \alpha \equiv |\alpha| e^{i\theta} \text{ and } \zeta_k = \zeta_{-k} = \zeta \equiv \xi e^{i\phi}. \tag{5.7}$$

As it was shown in [17], in this case, the state of the output light is completely separable from the state of the absorber though the reasons of this effect were not explained. The states of the input modes *k* and *-k* are depicted by 'coherent' arrows in a phase space with coordinates

$$\langle \hat{X}_{k,1}^{(in)} \rangle = \langle \hat{X}_{-k,1}^{(in)} \rangle = 2|\alpha|(\cos\theta \cosh\xi - \cos(\theta - \phi) \sinh\xi),$$

$$\langle \hat{X}_{k,2}^{(in)} \rangle = \langle \hat{X}_{-k,1}^{(in)} \rangle = 2|\alpha|(\sin\theta \cosh\xi + \sin(\theta - \phi) \sinh\xi),$$

and variance

$$\left(\Delta X_{k,1}^{(in)}\right)^2 = \left(\Delta X_{-k,1}^{(in)}\right)^2 = \cosh 2\xi - \cos\phi \sinh 2\xi, \tag{5.8}$$

$$\left(\Delta X_{k,2}^{(in)}\right)^2 = \left(\Delta X_{-k,2}^{(in)}\right)^2 = \cosh 2\xi + \cos\phi \sinh 2\xi. \tag{5.9}$$

For instance, for $\theta = \phi = 0$, each state is represented by the horizontal 'coherent' arrow with the uncertainty ellipse of radii $e^{-\xi}$ and $e^{\xi}$, as it is schematically shown in '*Input state*' box in Fig. 4 (top red and bottom blue states). Variance of the relative position (5.2) and total momentum (5.3) of the input modes *k* and *-k*,

$$\left(\Delta Q_{k,-k}^{\parallel}\right)^2 = \cosh 2\xi - \cos\phi \sinh 2\xi, \tag{5.10}$$

$$\left(\Delta P_{k,-k}^{\parallel}\right)^2 = \cosh 2\xi + \cos\phi \sinh 2\xi, \tag{5.11}$$

indicate lack of entanglement,

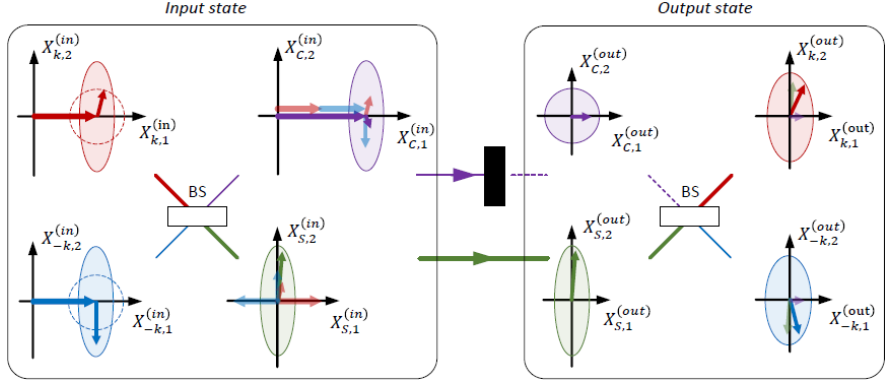

FIG. 4. **CPA of identically squeezed states**. From left to right: Identically squeezed states (top red and bottom blue, coherent state uncertainty is shown for reference by dashed circles) of the input modes *k* and *-k* are transformed into squeezed states (top purple and bottom green ellipses) of the standing waves by the first BS-transformation. The cosine standing wave is absorbed and its state is replaced by a vacuum state (top purple circle) while the sine wave is not affected. The state of the output modes *k* and *-k* (top red and bottom blue ellipses) is defined by the second BS-transformation. Each state is represented by its 'coherent' component (thick arrow) and uncertainty region. Randomly chosen thin arrow within uncertainty region corresponds to a particular realisation of the state. States evolution symbolically shown by transformation of arrows and uncertainty regions.

$$S_{k,-k}^{\|} = e^{2\xi} + e^{-2\xi} \geq 2, \quad (5.12)$$

as it should be for the input separable state (index '$\|$' stands for parallel orientation of the squeezing ellipses of two identical input states). The first BS-transformation results in squeezed states of the standing waves,

$$\left(\Delta X_{C,1}\right)^2 = \left(\Delta X_{S,1}\right)^2 = \left(\Delta X_{k,1}^{(in)}\right)^2,$$

$$\left(\Delta X_{C,2}\right)^2 = \left(\Delta X_{S,2}\right)^2 = \left(\Delta X_{k,2}^{(in)}\right)^2,$$

where the in-phase 'coherent' components are transferred to the cosine wave leaving the sine wave in a squeezed vacuum state (purple and green states in '*Input state*' box in Fig. 4):

$$\langle \hat{X}_{C,1}^{(in)} \rangle = \sqrt{2} \langle \hat{X}_{k,1}^{(in)} \rangle, \quad \langle \hat{X}_{C,2}^{(in)} \rangle = \sqrt{2} \langle \hat{X}_{k,2}^{(in)} \rangle, \quad (5.13)$$

$$\langle \hat{X}_{S,1}^{(in)} \rangle = \langle \hat{X}_{S,2}^{(in)} \rangle = 0. \quad (5.14)$$

Thus, correlations between the relative position $\hat{Q}_{C,S} = \left(\hat{X}_{C,1}^{(in)} - \hat{X}_{S,1}^{(in)}\right)/\sqrt{2}$ and total momentum $\hat{P}_{C,S} = \left(\hat{X}_{C,2}^{(in)} + \hat{X}_{S,2}^{(in)}\right)/\sqrt{2}$ of the standing waves repeat of those of the input travelling waves, and

$$S_{C,S}^{\|} = S_{k,-k}^{\|} = e^{2\xi} + e^{-2\xi} \geq 2,$$

where $S_{C,S}^{\|}$ is inseparability parameter of the standing waves. Since the cosine wave is imprinted to the absorber, the output light-absorber state will be separable as well. Moreover, this result stays valid for $\alpha_k \neq \alpha_{-k}$ in (5.7) since (5.8)-(5.12) are independent from these values. Thus, separability of the light-absorber state arises from the fact that two identically squeezed states do not produce entanglement through a BS-transformation. Next, according to (5.13) and (5.14) the regime of total coherence absorption takes place for the state parameters (5.7). At the same time, intensity is not fully absorbed, since the noise components of the input travelling waves are equally distributed between the standing waves. For the same reason, in general, the regimes of total absorption and total transmission of intensity are not possible for any squeezed (as well as for entangled) states. Finally, the state of the output modes *k* and *-k* can be restored as a result of the BS-transformation (5.1) of squeezed and coherent vacuum states ('*Output state*' box in Fig. 4) with

$$\langle \hat{X}_{k,i}^{(out)} \rangle = \langle \hat{X}_{-k,i}^{(out)} \rangle = 0,$$

and

$$\left(\Delta X_{k,1}^{(out)}\right)^2 = \left(\Delta X_{-k,1}^{(out)}\right)^2 = \frac{(1 + \cosh 2\xi - \cos\phi \sinh 2\xi)}{2},$$

$$\left(\Delta X_{k,2}^{(out)}\right)^2 = \left(\Delta X_{-k,2}^{(out)}\right)^2 = \frac{(1 + \cosh 2\xi + \cos\phi \sinh 2\xi)}{2}.$$

that is as a non-ideal squeezed vacuum states with $\left(\Delta X_{k,1}^{(out)}\right)^2 \left(\Delta X_{k,2}^{(out)}\right)^2 = \left(\Delta X_{-k,1}^{(out)}\right)^2 \left(\Delta X_{-k,2}^{(out)}\right)^2 > 1$.

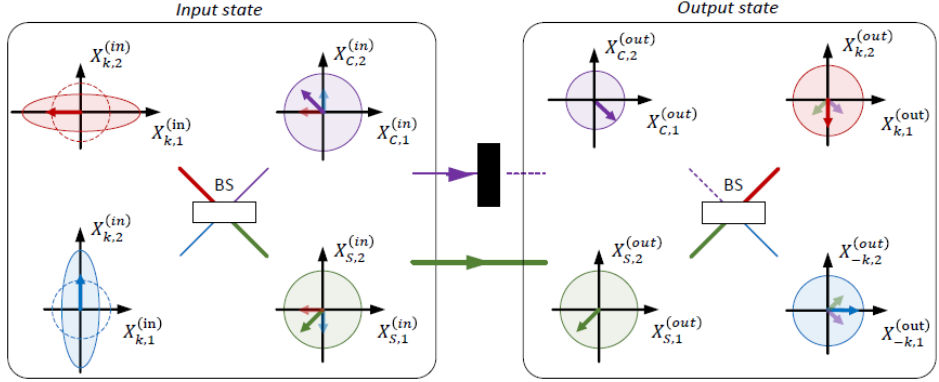

FIG. 5. **CPA of orthogonally squeezed states**. From left to right: Orthogonally squeezed vacuum states (top red and bottom blue ellipses) of the input modes *k* and -*k* are transformed into entangled states (top purple and bottom green circles of large radius) of the standing waves by the first BS-transformation. The cosine standing wave is fully absorbed and its state is replaced by vacuum while the sine wave is not affected. State of the output modes *k* and -*k* (top red and bottom blue circles) is defined by the second BS-transformation. Also, see caption to Fig. 4.

The opposite outcome – maximally entangled absorber-light state – would be of great interest for the protocols of light-matter interaction [21,78]. To achieve this, we use the fact that a BS-transformation generates entanglement given two input modes are in orthogonally squeezed states. Let us consider two input travelling modes in the states $|0, -\xi\rangle_k$ and $|0, \xi\rangle_{-k}$ with real $\xi > 0$ (that is $\phi_k = \pi$ and $\phi_{-k} = 0$). These states are shown in '*Input state*' box in Fig. 5. Compared to (5.8) and (5.9), quadratures of two input modes are squeezed along different directions:

$$\left(\Delta X_{k,1}^{(in)}\right)^2 = \left(\Delta X_{-k,2}^{(in)}\right)^2 = e^{2\xi},$$

$$\left(\Delta X_{k,2}^{(in)}\right)^2 = \left(\Delta X_{-k,1}^{(in)}\right)^2 = e^{-2\xi}.$$

As it is expected for the input separable state, no correlations are presented between the modes:

$$S_{k,-k}^{\perp} = e^{2\xi} + e^{-2\xi} > 2,$$

(index '⊥' stands for orthogonal orientation of the squeezing ellipses of two input states). After the first BS-transformation, we get the standing waves with 'noisy' quadratures (purple and green states in '*Input state*' box in Fig. 5),

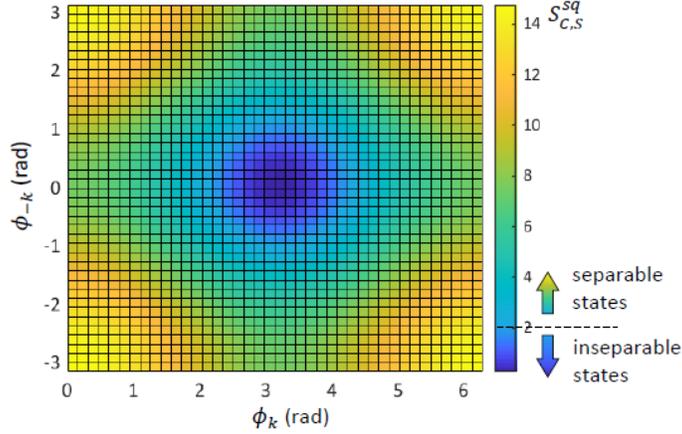

FIG. 6. **Inseparability parameter of the standing waves state for CPA of squeezed states**. Inseparability parameter $S_{C,S}^{sq}$ as a function of squeezing angles $\phi_k$ and $\phi_{-k}$ of the input travelling waves for equal squeezing parameters $\xi_k = \xi_{-k} = 1$. Black-to-blue region corresponds to inseparable, or entangled state of the standing waves, while the rest area corresponds to separable state. Since the cosine wave is imprinted to the absorber, the same graph represents inseparability parameter for the output light-absorber state.

$$\left(\Delta X_{C,1}^{(in)}\right)^2 = \left(\Delta X_{C,2}^{(in)}\right)^2 = \left(\Delta X_{S,1}^{(in)}\right)^2 = \left(\Delta X_{S,2}^{(in)}\right)^2 = \frac{e^{2\xi}+e^{-2\xi}}{2},$$

but their relative position and total momentum (defined similar to (5.2) and (5.3)), exhibit strong correlations:

$$S_{C,S}^\perp = 2e^{-2\xi} < 2. \tag{5.15}$$

Entangled state of the standing waves is then transferred to the light-absorber system. This result holds valid if the input states are built from non-zero coherent components $\alpha_k$ and $\alpha_{-k}$. The second BS-transformation is adverse for entanglement (due to amplitude splitting and addition of the vacuum noise) but can be reversed by sending the output modes $k$ and $-k$ through a 'real' 50:50 beamsplitter. One of the outputs of the 'real' beamsplitter recovers the sine wave amplitude,

$$\frac{1}{\sqrt{2}}\left(\hat{a}_k^{(out)} - \hat{a}_{-k}^{(out)}\right).$$

since two BS-transformations are equivalent to identity matrix.

For a general input states $|\alpha_k,\zeta_k\rangle_k$ and $|\alpha_{-k},\zeta_{-k}\rangle_{-k}$ with arbitrary parameters, inseparability is a function of ellipses orientation and squeezing parameters:

$$S_{C,S}^{sq} = \cosh 2\xi_k + \cosh 2\xi_{-k} + \cos\phi_k \sinh 2\xi_k - \cos\phi_{-k} \sinh 2\xi_{-k}.$$

For equal squeezing parameters, $\xi_k = \xi_{-k} = \xi$, inseparability reaches minimum at $\phi_k = \pi$ and $\phi_{-k} = 0$, as it is expected for orthogonally squeezed states (5.15), Fig. 6. The state of the standing waves is still inseparable within the black-to-blue area in Fig. 6, radius of which depends on the squeezing parameter $\xi$. Outside of this region, the states are separable. The latter regime is a generalization of the condition (5.7). For fixed angles $\phi_k = \pi$ and $\phi_{-k} = 0$ but $\xi_1 \neq \xi_2$, inseparability parameter for the standing waves decreases with an increase of the squeezing parameters:

$$S_{C,S}^{Sq} = e^{-2\xi_1} + e^{-2\xi_2} < 2.$$

As we have shown, the standing wave approach allows not only to determine absorption of light intensity and coherence, but also to follow quantum states evolution, providing a clear picture of the underlying processes.

## 5.3 CPA of CV entangled (EPR) states

In this section, we consider CV entangled, or EPR (Einstein-Podolsky-Rosen) state for input counter-propagating waves, $\hat{a}_k^{(in)}$ and $\hat{a}_{-k}^{(in)}$. As it was discussed in the previous section, entangled state arises from interference of orthogonally squeezed states on a 50:50 beamsplitter [57,79]. (Other ways to generate entangled states include parametric down-conversion process in non-linear crystals [80-82] and atomic ensembles [83-85].) Thus, it is beneficial, without loss of generality, to introduce two virtual 'preceding' modes $g$ and $h$ with corresponding amplitudes

$$\hat{g} = \frac{(\hat{X}_{g,1} + i\hat{X}_{g,2})}{2}, \quad \hat{h} = \frac{(\hat{X}_{h,1} + i\hat{X}_{h,2})}{2},$$

which are prepared in orthogonally squeezed states $|\alpha_g, -\xi\rangle_g$ and $|\alpha_h, \xi\rangle_h$. Here $\alpha_g = |\alpha_g|e^{i\theta_g}$ and $\alpha_h = |\alpha_h|e^{i\theta_h}$ are complex numbers, and squeezing parameter $\xi > 0$ is a real number. Average values of $\hat{g}$ and $\hat{h}$ [26],

$$\langle \hat{g} \rangle = e^{\xi} Re(\alpha_g) + ie^{-\xi} Im(\alpha_g) \equiv \alpha'_g, \tag{5.16}$$

$$\langle \hat{h} \rangle = e^{-\xi} Re(\alpha_h) + ie^{\xi} Im(\alpha_h) \equiv \alpha'_h, \tag{5.17}$$

define coordinates of 'coherent' arrows in a phase space with variance

$$(\Delta X_{g,1})^2 = (\Delta X_{h,2})^2 = e^{2\xi},$$

$$(\Delta X_{g,2})^2 = (\Delta X_{h,1})^2 = e^{-2\xi}.$$

The input travelling waves of interest, $\hat{a}_k^{(in)}$ and $\hat{a}_{-k}^{(in)}$, are considered as a result of BS-transformation of the 'preceding' modes:

$$\begin{pmatrix} \hat{a}_k^{(in)} \\ \hat{a}_{-k}^{(in)} \end{pmatrix} = H_{BS} \begin{pmatrix} \hat{g} \\ \hat{h} \end{pmatrix}$$

with matrix (2.11). The state of modes $k$ and $-k$ are reflected in a phase space by using their average values,

$$\begin{pmatrix} \langle \hat{a}_k^{(in)} \rangle \\ \langle \hat{a}_{-k}^{(in)} \rangle \end{pmatrix} = H_{BS} \begin{pmatrix} \alpha'_g \\ \alpha'_h \end{pmatrix} \equiv \begin{pmatrix} \alpha_k \\ \alpha_{-k} \end{pmatrix}, \tag{5.18}$$

and variance

$$(\Delta X_{k,1})^2 = (\Delta X_{k,2})^2 = (\Delta X_{-k,1})^2 = (\Delta X_{-k,2})^2 = \frac{e^{2\xi} + e^{-2\xi}}{2}.$$

Large uncertainty regions of the individual modes accompanied by strong correlations in the relative position and total momentum is characteristic of entangled states, and, indeed,

$$(\Delta Q_{k,-k})^2 = (\Delta P_{k,-k})^2 = e^{-2\xi},$$

$$S_{k,-k}^{Ent} = 2e^{-2\xi} < 2.$$

The greater 'preceding' squeezing, the more entangled states one gets. Entangled state of the input travelling waves is shown by red and blue circles in '*Input state*' box in Fig. 7 for $\alpha_k =$ $\alpha_{-k} = 0$. Transformation of the counter-propagating waves into the standing waves results in a simple outcome – the standing waves repeat the 'preceding' modes:

$$\begin{pmatrix} \hat{A}_C^{(in)} \\ \hat{A}_S^{(in)} \end{pmatrix} = H_{BS} \begin{pmatrix} \hat{a}_k^{(in)} \\ \hat{a}_{-k}^{(in)} \end{pmatrix} = I \begin{pmatrix} \hat{g} \\ \hat{h} \end{pmatrix},$$

where *I* is an identity matrix. The cosine wave is excited into the squeezed state $|\alpha_g, -\xi\rangle_C$ with the well-defined momentum

$$\left(\Delta X_{C,2}^{(in)}\right)^2 = e^{-2\xi},$$

and the sine wave is excited into the squeezed state $|\alpha_h, \xi\rangle_S$ with the well-defined position

$$\left(\Delta X_{S,1}^{(in)}\right)^2 = e^{-2\xi}.$$

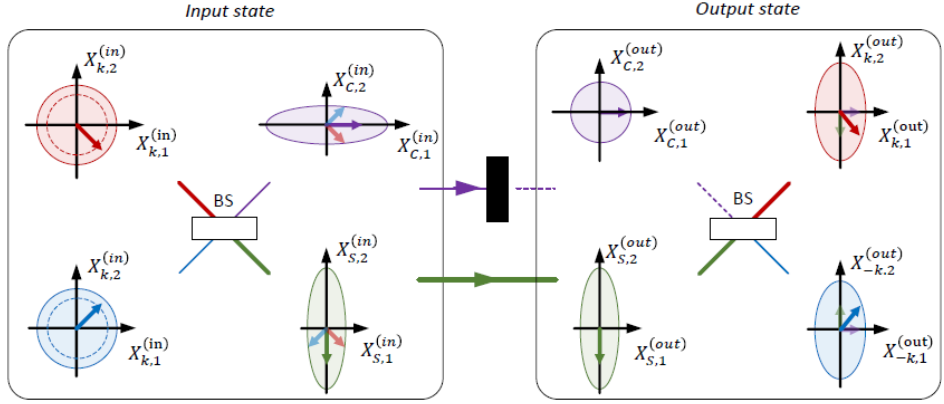

FIG. 7. **CPA of CV entangled (EPR) states.** From left to right: Entangled state (top red and bottom blue circles of large radius) of the input modes *k* and *-k* is transformed into orthogonally squeezed states (top purple and bottom green ellipses) of the standing waves by the first BS-transformation. The cosine standing wave is fully absorbed and its state is replaced by vacuum while the sine standing wave is not affected. State of the output modes *k* and *-k* (top red and bottom blue) is defined by the second BS-transformation. Also, see caption to Fig. 4.

Since the standing waves state is separable,

$$S_{C,S}^{Ent} = e^{2\xi} + e^{-2\xi} > 2,$$

the state of the absorber-output light system will be separable as well. Intensity, presented initially in the mode *g*,

$$\langle \hat{g}^\dagger \hat{g} \rangle = |\alpha_g|^2 \cosh 2\xi + |\alpha_g|^2 \cos 2\theta_g \sinh 2\xi + \sinh^2 \xi, \qquad (5.19)$$

is absorbed, while intensity of the mode *h*,

$$\langle \hat{h}^\dagger \hat{h} \rangle = |\alpha_h|^2 \cosh 2\xi - |\alpha_h|^2 \cos 2\theta_h \sinh 2\xi + \sinh^2 \xi, \qquad (5.20)$$

is split between the output light modes. Intensity absorption coefficient can be found as

$$\mathcal{A}_{ent}^I = \frac{\langle \hat{g}^\dagger \hat{g} \rangle}{\langle \hat{g}^\dagger \hat{g} \rangle + \langle \hat{h}^\dagger \hat{h} \rangle}, \qquad (5.21)$$

that is as a function of parameters of the 'preceding' modes *g* and *h*. By reversing the transformation (5.18) and using (5.16) and (5.17), we may express $\mathcal{A}_{ent}^I$ as a function of the parameters of the input modes *k* and *-k*. For instance, for $|\alpha_k| = |\alpha|e^{i\theta_k}$ and $|\alpha_{-k}| = |\alpha|e^{i\theta_{-k}}$, we get:

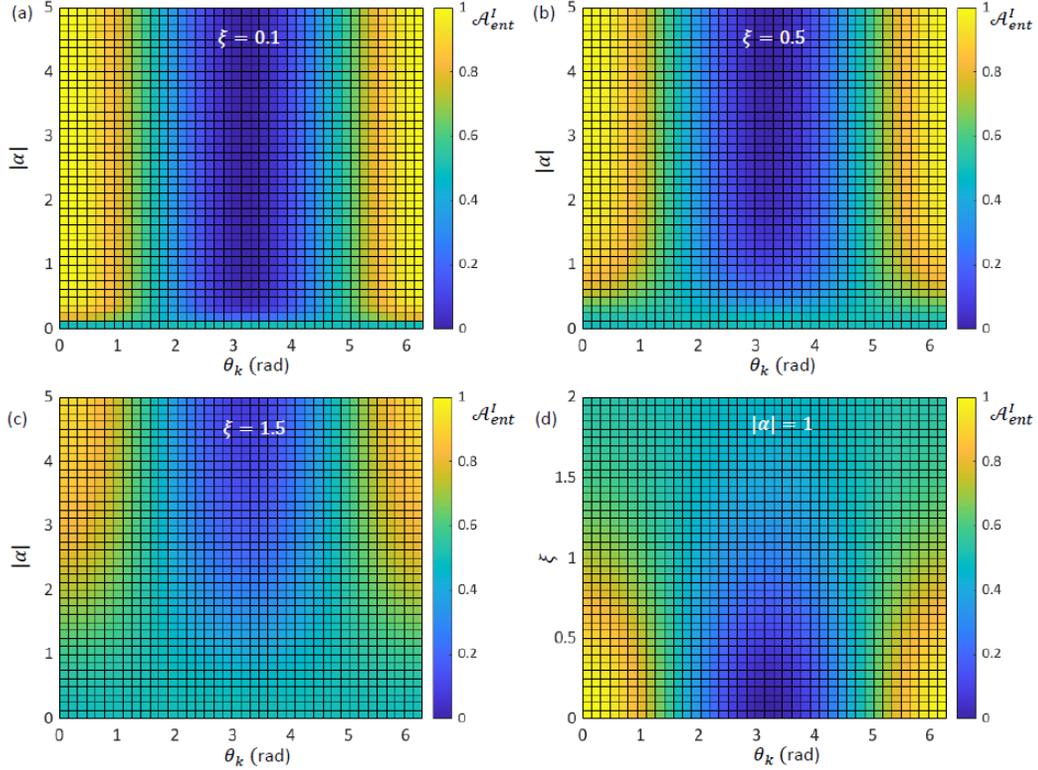

FIG. 8. **Intensity absorption of CV entangled (EPR) states.** (a)-(c) Intensity absorption coefficient $\mathcal{A}_{ent}^I$ as a function of the phase $\theta_k$ and amplitude $|\alpha|$ of the input mode $k$. Squeezing parameter $\xi$ of 'preceding' modes equals to 0.1 (a), 0.5 (b) and 1.5 (c). The input travelling waves are assumed to have equal coherent amplitudes $|\alpha_k| = |\alpha_{-k}| = |\alpha|$ and angle $\theta_{-k}$ is set to zero. (d) $\mathcal{A}_{ent}^I$ as a function of the phase $\theta_k$ and squeezing parameter $\xi$ with fixed coherent components, $|\alpha_k| = |\alpha_{-k}| = 1$.

$$\theta_g = \mathrm{atan}\left[e^{2\xi}\tan\frac{\theta_k + \theta_{-k}}{2}\right],$$

$$\theta_h = \mathrm{atan}\left[-e^{-2\xi}\cot\frac{\theta_k + \theta_{-k}}{2}\right],$$

$$|\alpha_g| = \sqrt{2}e^{-\xi}|\alpha|\cos\frac{\Delta\theta}{2}\cos\frac{\theta_k + \theta_{-k}}{2}/\cos\theta_g,$$

$$|\alpha_h| = -\sqrt{2}e^{\xi}|\alpha|\sin\frac{\Delta\theta}{2}\sin\frac{\theta_k + \theta_{-k}}{2}/\cos\theta_h,$$

where $\Delta\theta = \theta_k - \theta_{-k}$. These relations should be substituted into (5.19) and (5.20) and, finally, to (5.21). Intensity absorption coefficient as a function of $\theta_k$ and $|\alpha|$ ($\theta_{-k} = 0$) for different values of $\xi$ is shown in Figs. 8(a)-(c). For small $\xi$, Fig. 8(a), the input modes state is close to coherent ('classical') state with $S_{k,-k}^{ent} \approx 2$. In this case, close to classical regime behaviour is observed with corresponding oscillation between the regimes of total absorption and total transmission independent on coherent amplitude of input light $|\alpha|$. With increase of $\xi$, Fig. 8(b) and 8(c), inseparability parameter decreases and non-classical behaviour takes place, where absorption depends on both the phase relation between the input modes and the ratio between $|\alpha|$ and $\xi$. When intensity of coherent component of input light, $|\alpha|^2$, is greater than intensity of incoherent component (intensity of fluctuations, $\sinh^2\xi$) classical behaviour dominates. In opposite case of large incoherent component, light is split equally between the cosine and sine standing waves with following $\approx 50\%$ phase-independent absorption. This transition from the classical to quantum regimes with the increase of squeezing parameter $\xi$ is clearly seen in Fig. 8(d) where coherent amplitude is fixed, $|\alpha| = 1$. In the case of vacuum entangled state, $\alpha = 0$, only random fluctuation component is presented, and $\mathcal{A}^I_{ent} = 1/2$ for any degree of entanglement.

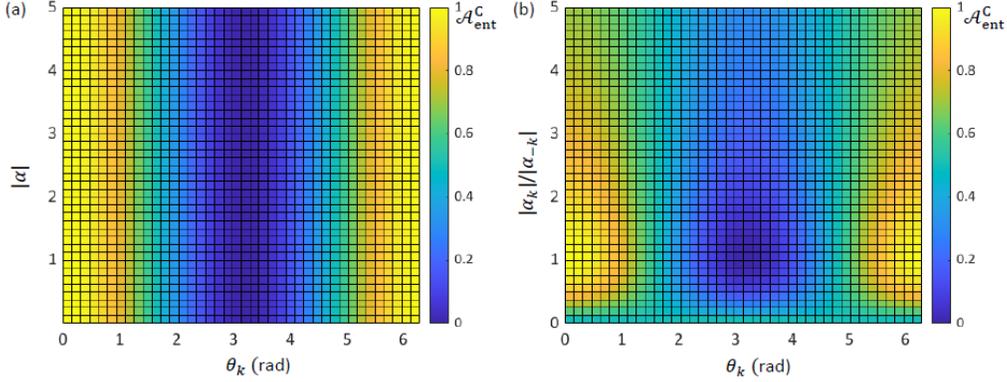

FIG. 9. **Coherence absorption of CV entangled (EPR) states**. (a) Coherence absorption coefficient $\mathcal{A}^C_{ent}$ as a function of the phase $\theta_k$ and amplitude $|\alpha|$ of the input mode $k$. The input travelling waves are assumed to have equal coherent amplitudes $|\alpha_k| = |\alpha_{-k}| = |\alpha|$ and the phase $\theta_{-k}$ is set to zero. (b) $\mathcal{A}^C_{ent}$ as a function of the phase $\theta_k$ and ratio $|\alpha_k|/|\alpha_{-k}|$. In both cases $\theta_{-k} = 0$.

Similar, coherence of the 'preceding' mode $g$, $|\langle \hat{g} \rangle|^2 = |\alpha'_g|^2$, is absorbed, while coherence of the mode $h$, $|\langle \hat{h} \rangle|^2 = |\alpha'_h|^2$, is transmitted to the output light modes. Coherence absorption coefficient (5.6),

$$\mathcal{A}^C_{ent} = \frac{|\langle \hat{g} \rangle|^2}{|\langle \hat{g} \rangle|^2 + |\langle \hat{h} \rangle|^2},$$

is expressed through parameters of the input modes $k$ and $-k$ by using transformation (5.18):

$$\mathcal{A}^C_{ent} = \frac{|\alpha_k|^2 + |\alpha_{-k}|^2 + 2|\alpha_k||\alpha_{-k}|\cos \Delta\theta}{2(|\alpha_k|^2 + |\alpha_{-k}|^2)},$$

For $|\alpha_k| = |\alpha_{-k}|$, coherence absorption coefficient

$$\mathcal{A}^C_{ent} = \frac{1 + \cos \Delta\theta}{2}, \tag{5.22}$$

is a function of the phase difference $\Delta\theta$ between coherent components of the corresponding amplitudes (or the angle between 'coherent' arrows in a phase space). As we discussed previously, coherence defines component of light which interfere in a classical sense, and equation (5.22) reflects the law of the classical regime of CPA. If all coherence of the input light is transferred to the cosine standing wave ($\Delta\theta = 2\pi n$), the regime of total absorption of coherence follows. Opposite, if all coherence is transferred to the sine standing wave ($\Delta\theta = (2n+1)\pi$), the regime of total transmission of coherence takes place. To compare absorption of coherence with absorption of intensity, we plot $\mathcal{A}^C_{ent}$, Fig. 9(a), for the same set of parameters as it is done for $\mathcal{A}^I_{ent}$ in Fig. 8(a)-(c). For small squeezing parameters, absorption of intensity, Fig. 8(a), is almost equivalent to absorption of coherence. In this case, the state of input light is close to coherent (classical) state, and coherence and intensity is, basically, the same variable. For instance, for input mode $k$ one gets $I_k \approx |\langle \hat{a}^{(in)}_k \rangle|^2 = C_k$. By increasing the parameter $\xi$ (equivalent to increasing the degree of entanglement of the input counter-

propagating modes), one increases the amount of energy presented in random fluctuations and intensity absorption coefficient tends to 1/2. At the same time, coherence absorption coefficient, independent of $\xi$, still oscillates between 0 and 1. Thus, absorption of coherence can either be greater or less than absorption of intensity. For $|\alpha_k| \neq |\alpha_{-k}|$, coherence absorption coefficient tends to 1/2 with increase of $|\alpha_k|/|\alpha_{-k}|$ ratio, Fig. 9(b).

*5.4 CPA of other CV states*

In a similar manner, interference of other CV quantum states may be considered. As a first example, we shortly outline CPA of Schrödinger cat states (SCS). In this case, the input counter-propagating modes $k$ and $-k$ are set in a superposition of coherent states of opposite phases [86,87] (normalization factor is omitted):

$$|SC(\alpha)\rangle_k \sim |\alpha\rangle_k + |-\alpha\rangle_k,$$

$$|SC(\alpha)\rangle_{-k} \sim |\alpha\rangle_{-k} + |-\alpha\rangle_{-k}.$$

The first BS-transformation results in a superposition of Schrödinger cat states of the standing waves of the form [88-90],

$$|SC(\sqrt{2}\alpha)\rangle_C |0\rangle_S + |0\rangle_C |SC(\sqrt{2}\alpha)\rangle_S.$$

Thus, all input light is absorbed with probability of 50%. With the same probability of 50%, all light is transmitted and the state of the output light modes can be found as

$$|SC(\sqrt{2}\alpha)\rangle_S \to |\alpha\rangle_k |-\alpha\rangle_{-k} + |-\alpha\rangle_k |\alpha\rangle_{-k}.$$

As a second example, we consider the regime of 'asymmetric' illumination of the absorber by coherent $|\alpha\rangle_k$ and squeezed $|0,\zeta\rangle_{-k}$ states. The first BS-transformation generates *NOON* state of the standing waves [91,92]:

$$\frac{1}{\sqrt{2}}(|N\rangle_C |0\rangle_S + |0\rangle_C |N\rangle_S),$$

where the number of excitations is defined by the ratio $|\alpha|^2/\xi$. Again, light is absorbed or transmitted with 50% probability.

In a third example, we consider 'asymmetric' illumination of the absorber by coherent $|\alpha\rangle_k$ and Schrödinger cat $|\alpha\rangle_{-k} + |-\alpha\rangle_{-k}$ states [93]. The first BS-transformation results in the entangled state of the standing waves

$$\sim |\sqrt{2}\alpha\rangle_C |0\rangle_S + |0\rangle_C |\sqrt{2}\alpha\rangle_S,$$

where the input light can be found either in the cosine or the sine standing waves. For all three cases considered here $\langle \hat{a}_k^{(in)} \rangle^* \langle \hat{a}_{-k}^{(in)} \rangle = 0$ and intensity and coherence absorption coefficients are equal to 1/2.

In a similar manner, analysis of CPA of other CV quantum states can be done, which we leave out of the scope of this paper.

## 6. Summary and conclusions

We shortly summarize the discussed above quantum regimes of CPA in Table 1. Single photon state absorption follows the classical regime of CPA where the light absorption may be manipulated in the full range between the total absorption and total transmission. Quantum fluctuations presented in other discrete variable states do not allow to manipulate the average light absorption which is always 50% since light is distributed equally between the standing waves. Meanwhile, the number of photons dissipated in a single act of absorption differs drastically depending on the particular input state of light. This allows to implement absorption of a given Fock state with a high probability while dumping, for the certain extent, absorption of other Fock states. Similar situation is observed for continuous variable quantum states where quantum fluctuations are evenly spread between the standing waves preventing manipulation of the average light absorption probability. At the same time, when the state of light approaches 'classical' coherent state (for instance by setting $\xi \to 0$ for squeezed and entangled states) the absorption becomes a phase sensitive and follows classical patterns. Again, individual events of absorption differ significantly from equally transmitted and dissipated light intensities to probabilistic total absorption or total transmission of input light intensity depending on the input state of light. Separability of the output light-absorber state is defined by the separability of the standing waves state.

Table 1. Coherent perfect absorption of quantum states of light.

| | State type | Input state (Travelling waves) | Standing waves state | Average intensity absorption | Notes |
|---|---|---|---|---|---|
| **Discrete variable states** | Single photon — Superposition state | $\frac{1}{\sqrt{2}}(|1\rangle_k|0\rangle_{-k} + e^{i\Delta\theta}|0\rangle_k|1\rangle_{-k})$ | $\cos\frac{\Delta\theta}{2}|1\rangle_C|0\rangle_S - i\sin\frac{\Delta\theta}{2}|0\rangle_C|1\rangle_S$ | $\cos^2\frac{\Delta\theta}{2}$ | Controllable absorption: from total absorption to total transmission |
| | Two photon states — Triplet Bell state (bosonic) | $|\psi^{(+)}\rangle_{k,-k} = \frac{1}{\sqrt{2}}(|A\rangle_k|B\rangle_{-k} + |B\rangle_k|A\rangle_{-k})$ | $|\varphi^{(-)}\rangle_{C,S} = \frac{1}{\sqrt{2}}(|A\rangle_C|B\rangle_C - |A\rangle_S|B\rangle_S)$ | 50% | 2-photon absorption (50%) 0-photon absorption (50%) |
| | Two photon states — Singlet Bell state (fermionic) | $|\psi^{(-)}\rangle_{k,-k} = \frac{1}{\sqrt{2}}(|A\rangle_k|B\rangle_{-k} - |B\rangle_k|A\rangle_{-k})$ | $|\psi^{(-)}\rangle_{C,S} = \frac{1}{\sqrt{2}}(|A\rangle_C|B\rangle_S - |B\rangle_C|A\rangle_S)$ | 50% | 1-photon absorption (100%) |
| | Two photon states — Triplet Bell state (NOON+) | $|\varphi^{(+)}\rangle_{k,-k} = \frac{1}{\sqrt{2}}(|A\rangle_k|B\rangle_{-k} + |A\rangle_{-k}|B\rangle_{-k})$ | $|\varphi^{(+)}\rangle_{C,S} = \frac{1}{\sqrt{2}}(|A\rangle_C|B\rangle_C + |A\rangle_S|B\rangle_S)$ | 50% | 2-photon absorption (50%) 0-photon absorption (50%) |
| | Two photon states — Triplet Bell state (NOON-) | $|\varphi^{(-)}\rangle_{k,-k} = \frac{1}{\sqrt{2}}(|A\rangle_k|B\rangle_k - |A\rangle_{-k}|B\rangle_{-k})$ | $|\psi^{(+)}\rangle_{C,S} = \frac{1}{\sqrt{2}}(|A\rangle_C|B\rangle_S + |B\rangle_C|A\rangle_S)$ | 50% | 1-photon absorption (100%) |
| | NOON states — NOON state with N=3 | $\frac{1}{\sqrt{2}}(|3\rangle_k|0\rangle_{-k} + e^{i\Delta\theta}|0\rangle_k|3\rangle_{-k})$ | $\frac{1}{2}\{\cos\frac{\Delta\theta}{2}(|3\rangle_C|0\rangle_S - \sqrt{3}|1\rangle_C|2\rangle_S) + \sin\frac{\Delta\theta}{2}(|0\rangle_C|3\rangle_S - \sqrt{3}|2\rangle_C|1\rangle_S)\}$ | 50% | $\Delta\theta = 2\pi n$: 3-photon absorption (25%), 1-photon absorption (75%); $\Delta\theta = (2n+1)\pi$: 2-photon absorption (75%), 0-photon absorption (25%) |
| | NOON states — NOON state with N=4 | $\frac{1}{\sqrt{2}}(|4\rangle_k|0\rangle_{-k} + e^{i\Delta\theta}|0\rangle_k|4\rangle_{-k})$ | $\frac{1}{2\sqrt{2}}\{\cos\frac{\Delta\theta}{2}(|4\rangle_C|0\rangle_S + |0\rangle_C|4\rangle_S - \sqrt{6}|2\rangle_C|2\rangle_S) + 2\sin\frac{\Delta\theta}{2}(|3\rangle_C|1\rangle_S - |1\rangle_C|3\rangle_S)\}$ | 50% | $\Delta\theta = 2\pi n$: 4-photon absorption (12.5%), 2-photon absorption (75%), 0-photon absorption (12.5%); $\Delta\theta = (2n+1)\pi$: 3-photon absorption (50%), 1-photon absorption (50%) |
| **Continuous variable states** | Squeezed states — Identical squeezed states | $|\alpha,\xi\rangle_k|\alpha,\xi\rangle_{-k}$ | $|\sqrt{2}\alpha,\xi\rangle_C|0,\xi\rangle_S$ | 100% for $\xi = 0$ 50% for $\xi \to \infty$ | *Separable* output light-absorber state |
| | Squeezed states — Orthogonally squeezed vacuum states | $\|0,-\xi\rangle_k\|0,\xi\rangle_{-k}$ Squeezing in orthogonal quadratures Separable state: $S_{k,-k}^\perp > 2$ | Individual quadratures are uncertain Inseparable (entangled) state: $S_{C,S}^\perp < 2$ | 50% | *Inseparable* output light-absorber state |
| | EPR state — Entangled state | Inseparable (entangled) state: $S_{k,-k}^{Ent} < 2$ | Separable squeezed states: $S_{C,S}^{Ent} > 2$ | 50% for ideally entangled states | *Separable* output light-absorber state |
| | Schrödinger cat states — Identical cat states ($\|SC(\alpha)\rangle \sim \|\alpha\rangle + \|-\alpha\rangle$) | $|SC(\alpha)\rangle_k|SC(\alpha)\rangle_{-k}$ | $|SC(\sqrt{2}\alpha)\rangle_C|0\rangle_S + |0\rangle_C|SC(\sqrt{2}\alpha)\rangle_S$ | 50% | All light absorbed (50%) All light transmitted (50%) *Inseparable* output light-absorber state |
| | Asymmetric input states — Coherent and squeezed states | $|\alpha\rangle_k|0,\zeta\rangle_{-k}$ | $\frac{1}{\sqrt{2}}(|N\rangle_C|0\rangle_S + |0\rangle_C|N\rangle_S)$ | 50% | All light absorbed (50%) All light transmitted (50%) *Inseparable* output light-absorber state |
| | Asymmetric input states — Coherent and Schrödinger cat states | $|\alpha\rangle_k|SC(\alpha)\rangle_{-k}$ | $|\sqrt{2}\alpha\rangle_C|0\rangle_S + |0\rangle_C|\sqrt{2}\alpha\rangle_S$ | 50% | All light absorbed (50%) All light transmitted (50%) *Inseparable* output light-absorber state |

In conclusion, we presented unified theory of quantum regime of CPA by considering the problem in the quantized standing waves basis taking account of a subwavelength thickness of the absorber. Our description allowed us to build a detailed picture of the underlying processes of CPA tracing evolution of quantum states of light as well as to bring strong parallels between classical and quantum regimes of CPA. Under a single theory, known quantum effects of CPA were explained and other regimes were introduced both for discrete and continuous variable quantum states. Deep understanding of fundamental properties of quantum light dissipation will allow practical implementation of this phenomenon in quantum optics and quantum information including manipulation and measurement of quantum states of light and excitation of non-classical states of light and matter.

The data that support the findings of this study are openly available in NTU research data repository DR-NTU (Data) at [94].

**Acknowledgments**

The authors are grateful to Nikolay I. Zheludev for fruitful discussion. This work was supported by the Singapore A*STAR QTE program (SERC A1685b0005), the Singapore Ministry of Education (MOE2016-T3-1-006 (S)) and Singapore National Research Foundation (NRF) Quantum Engineering Programme (QEP-P1).

**References**

[1] Y. D. Chong, L. Ge, H. Cao, and A. D. Stone, "Coherent perfect absorbers: time-reversed lasers," Phys. Rev. Lett. **105**, 053901 (2010).
[2] W. Wan, Y. D. Chong, L. Ge, H. Noh, A. D. Stone, and H. Cao, "Time-Reversed Lasing and Interferometric Control of Absorption," Science **331**, 889-892 (2011).
[3] D. G. Baranov, A. Krasnok, T. Shegai, A. Alu, and Y. D. Chong, "Coherent perfect absorbers: linear control of light with light," Nat. Rev. Mater. **2**, 17064 (2017).
[4] J. F. Zhang, K. F. MacDonald, and N. I. Zheludev, "Controlling light-with-light without nonlinearity," Light Sci. Appl. **1**, e18 (2012).
[5] X. Fang, M. L. Tseng, J. Y. Ou, K. F. MacDonald, D. P. Tsai, and N. I. Zheludev, "Ultrafast all-optical switching via coherent modulation of metamaterial absorption," Appl. Phys. Lett. **104**, 141102 (2014).
[6] X. Fang, K. F. MacDonald, and N. I. Zheludev, "Controlling light with light using coherent metadevices: all-optical transistor, summator and invertor," Light Sci. Appl. **4**, e292 (2015).
[7] A. Xomalis, I. Demirtzioglou, E. Plum, Y. M. Jung, V. Nalla, C. Lacava, K. F. MacDonald, P. Petropoulos, D. J. Richardson, and N. I. Zheludev, "Fibre-optic metadevice for all-optical signal modulation based on coherent absorption," Nat. Commun. **9**, 182 (2018).
[8] A. Xomalis, I. Demirtzioglou, Y. Jung, E. Plum, C. Lacava, P. Petropoulos, D. J. Richardson, and N. I. Zheludev, "Picosecond all-optical switching and dark pulse generation in a fibre-optic network using a plasmonic metamaterial absorber," Appl. Phys. Lett. **113**, 051103 (2018).
[9] A. Goodarzi, M. Ghanaatshoar, and M. Mozafari, "All-optical fiber optic coherent amplifier," Sci. Rep. **8**, 15340 (2018).
[10] T. Roger, S. Vezzoli, E. Bolduc, J. Valente, J. J. F. Heitz, J. Jeffers, C. Soci, J. Leach, C. Couteau, N. I. Zheludev, and D. Faccio, "Coherent perfect absorption in deeply subwavelength films in the single-photon regime," Nat. Commun. **6**, 7031 (2015).
[11] A. N. Vetlugin, R. Guo, A. Xomalis, S. Yanikgonul, G. Adamo, C. Soci, and N. I. Zheludev, "Coherent perfect absorption of single photons in a fiber network," Appl. Phys. Lett. **115**, 191101 (2019).
[12] T. Roger, S. Restuccia, A. Lyons, D. Giovannini, J. Romero, J. Jeffers, M. Padgett, and D. Faccio, "Coherent Absorption of N00N States," Phys. Rev. Lett. **117**, 023601 (2016).
[13] B. Vest, M. C. Dheur, E. Devaux, A. Baron, E. Rousseau, J. P. Hugonin, J. J. Greffet, G. Messin, and F. Marquier, "Anti-coalescence of bosons on a lossy beam splitter," Science **356**, 1373-1376 (2017).

[14] A. Lyons, D. Oren, T. Roger, V. Savinov, J. Valente, S. Vezzoli, N. I. Zheludev, M. Segev, and D. Faccio, "Coherent metamaterial absorption of two-photon states with 40% efficiency," Phys. Rev. A **99**, 011801 (2019).
[15] S. M. Barnett, J. Jeffers, A. Gatti, and R. Loudon, "Quantum optics of lossy beam splitters," Phys. Rev. A **57**, 2134-2145 (1998).
[16] J. Jeffers, "Interference and the lossless lossy beam splitter," J. Mod. Opt. **47**, 1819-1824 (2000).
[17] A. Ü. C. Hardal and M. Wubs, "Quantum coherent absorption of squeezed light," Optica **6**, 181-189 (2019).
[18] S. Yanikgonul, A. N. Vetlugin, R. Guo, A. Xomalis, G. Adamo, C. Soci, and N. I. Zheludev, "Quantum State Filtering of Dual-rail Photons with Fiberized Plasmonic Metamaterial", in Conference on Lasers and Electro-Optics, OSA Technical Digest (Optical Society of America, 2019), paper FTu3D.7.
[19] S. Huang and G. S. Agarwal, "Coherent perfect absorption of path entangled single photons," Opt. Express **22**, 20936-20947 (2014).
[20] A. N. Vetlugin, R. Guo, C. Soci, and N. I. Zheludev, "Anti-Hong-Ou-Mandel effect with entangled photons," arXiv:2105.05444 [quant-ph].
[21] A. N. Vetlugin, R. Guo, C. Soci, and N. I. Zheludev, "Deterministic Generation of Entanglement in Quantum Networks by Distributed Coherent Absorption," in *Conference on Lasers and Electro-Optics (CLEO)* (2021).
[22] S. Yanikgonul, R. Guo, A. Xomalis, A. N. Vetlugin, G. Adamo, C. Soci, and N. I. Zheludev, "Phase stabilization of a coherent fiber network by single-photon counting," Opt. Lett. **45**, 2740-2743 (2020).
[23] C. Altuzarra, S. Vezzoli, J. Valente, W. B. Gao, C. Soci, D. Faccio, and C. Couteau, "Coherent Perfect Absorption in Metamaterials with Entangled Photons," Acs Photonics **4**, 2124-2128 (2017).
[24] J. Y. Suen, K. Fan, and W. J. Padilla, "A Zero-Rank, Maximum Nullity Perfect Electromagnetic Wave Absorber," Adv. Opt. Mater. **7**, 1801632 (2019).
[25] R. Loudon, *The quantum theory of light* (Clarendon Press, Oxford, 1973).
[26] M. O. Scully and M. S. Zubairy, *Quantum Optics* (Cambridge University Press, Cambridge, 1997).
[27] A. Karvounis, V. Nalla, K. F. MacDonald, N. I. Zheludev, "Ultrafast Coherent Absorption in Diamond Metamaterials," Adv. Mater. **30**, e1707354 (2018)
[28] S. Thongrattanasiri, F. H. L. Koppens, and F. J. García de Abajo, "Complete Optical Absorption in Periodically Patterned Graphene," Phys. Rev. Lett. **108**, 047401 (2012).
[29] S. L. Braunstein and P. van Loock, "Quantum information with continuous variables," Rev. Mod. Phys. **77**, 513-577 (2005).
[30] C. Weedbrook, S. Pirandola, R. García-Patrón, N. J. Cerf, T. C. Ralph, J. H. Shapiro, and S. Lloyd, "Gaussian quantum information," Rev. Mod. Phys. **84**, 621-669 (2012).
[31] K. Mattle, H. Weinfurter, P. G. Kwiat, and A. Zeilinger, "Dense Coding in Experimental Quantum Communication," Phys. Rev. Lett. **76**, 4656-4659 (1996).
[32] Y. Guo, B.-H. Liu, C.-F. Li, and G.-C. Guo, "Advances in Quantum Dense Coding," Adv. Quantum Technol. **2**, 1900011 (2019).
[33] D. Bouwmeester, J.-W. Pan, K. Mattle, M. Eibl, H. Weinfurter, and A. Zeilinger, "Experimental quantum teleportation," Nature **390**, 575-579 (1997).
[34] S. Pirandola, J. Eisert, C. Weedbrook, A. Furusawa, and S. L. Braunstein, "Advances in quantum teleportation," Nat. Photonics **9**, 641-652 (2015).
[35] C. H. Bennett, G. Brassard, "Quantum cryptography: Public key distribution and coin tossing," Theor. Comput. Sci. **560**, 7-11 (2014).
[36] R. Bedington, J. M. Arrazola, and A. Ling, "Progress in satellite quantum key distribution," npj Quantum Inf. **3**, 30 (2017).
[37] M. A. Nielsen and I. L. Chuang, *Quantum Computation and Quantum Information*, 10th ed. (Cambridge University Press, Cambridge, 2010).
[38] T. Aoki, G. Takahashi, T. Kajiya, J.-i. Yoshikawa, S. L. Braunstein, P. van Loock, and A. Furusawa, "Quantum error correction beyond qubits," Nat. Physics **5**, 541-546 (2009).
[39] F. Arute, K. Arya, R. Babbush, D. Bacon, J. C. Bardin, R. Barends, R. Biswas, S. Boixo, F. G. S. L. Brandao, D. A. Buell, B. Burkett, Y. Chen, Z. Chen, B. Chiaro, R. Collins, W. Courtney, A. Dunsworth, E. Farhi, B. Foxen, A. Fowler, C. Gidney, M. Giustina, R. Graff, K. Guerin, S. Habegger, M. P. Harrigan, M. J. Hartmann, A. Ho, M. Hoffmann, T. Huang, T. S. Humble, S. V. Isakov, E. Jeffrey, Z. Jiang, D. Kafri, K. Kechedzhi, J. Kelly, P. V. Klimov, S. Knysh, A. Korotkov, F. Kostritsa, D. Landhuis, M. Lindmark, E. Lucero, D. Lyakh, S. Mandrà, J. R. McClean, M. McEwen, A. Megrant, X. Mi, K. Michielsen, M. Mohseni, J. Mutus, O. Naaman, M. Neeley, C. Neill, M. Y. Niu, E. Ostby, A. Petukhov, J. C. Platt, C. Quintana, E. G. Rieffel, P. Roushan, N. C. Rubin, D. Sank, K. J. Satzinger, V. Smelyanskiy, K. J. Sung, M. D. Trevithick, A. Vainsencher, B. Villalonga, T. White, Z. J. Yao, P. Yeh, A. Zalcman, H. Neven, and J. M. Martinis, "Quantum supremacy using a programmable superconducting processor," Nature **574**, 505-510 (2019).
[40] A. Nicolas, L. Veissier, L. Giner, E. Giacobino, D. Maxein, and J. Laurat, "A quantum memory for orbital angular momentum photonic qubits," Nat. Photonics **8**, 234-238 (2014).
[41] Y. Wang, J. Li, S. Zhang, K. Su, Y. Zhou, K. Liao, S. Du, H. Yan, and S.-L. Zhu, "Efficient quantum memory for single-photon polarization qubits," Nat. Photonics **13**, 346-351 (2019).

[42] R. Chrapkiewicz, M. Dąbrowski, and W. Wasilewski, "High-Capacity Angularly Multiplexed Holographic Memory Operating at the Single-Photon Level," Phys. Rev. Lett. **118**, 063603 (2017).
[43] S. A. Moiseev, K. I. Gerasimov, R. R. Latypov, N. S. Perminov, K. V. Petrovnin, and O. N. Sherstyukov, "Broadband multiresonator quantum memory-interface," Sci. Rep. **8**, 3982 (2018).
[44] P. Vernaz-Gris, K. Huang, M. Cao, A. S. Sheremet, and J. Laurat, "Highly-efficient quantum memory for polarization qubits in a spatially-multiplexed cold atomic ensemble," Nat. Commun. **9**, 363 (2018).
[45] M. Cao, F. Hoffet, S. Qiu, A. S. Sheremet, and J. Laurat, "Efficient reversible entanglement transfer between light and quantum memories," Optica **7**, 1440-1444 (2020).
[46] D. A. Kalashnikov, A. V. Paterova, S. P. Kulik, and L. A. Krivitsky, "Infrared spectroscopy with visible light," Nat. Photonics **10**, 98-101 (2016).
[47] J. Wang, F. Sciarrino, A. Laing, and M. G. Thompson, "Integrated photonic quantum technologies," Nat. Photonics **14**, 273-284 (2020).
[48] D. Llewellyn, Y. Ding, I. I. Faruque, S. Paesani, D. Bacco, R. Santagati, Y.-J. Qian, Y. Li, Y.-F. Xiao, M. Huber, M. Malik, G. F. Sinclair, X. Zhou, K. Rottwitt, J. L. O'Brien, J. G. Rarity, Q. Gong, L. K. Oxenlowe, J. Wang, and M. G. Thompson, "Chip-to-chip quantum teleportation and multi-photon entanglement in silicon," Nat. Physics **16**, 148-153 (2020).
[49] K. F. Lee, Y. Tian, H. Yang, K. Mustonen, A. Martinez, Q. Dai, E. I. Kauppinen, J. Malowicki, P. Kumar, Z. Sun, "Photon-Pair Generation with a 100 nm Thick Carbon Nanotube Film," Adv. Mater. **29**, 1605978 (2017).
[50] A. Zeilinger, H. J. Bernstein, and M. A. Horne, "Information Transfer with Two-state Two-particle Quantum Systems," J. Mod. Opt. **41**, 2375-2384 (1994).
[51] C. K. Hong, Z. Y. Ou, and L. Mandel, "Measurement of subpicosecond time intervals between two photons by interference," Phys. Rev. Lett. **59**, 2044-2046 (1987).
[52] J. Dunningham and T. Kim, "Using quantum interferometers to make measurements at the Heisenberg limit," J. Mod. Opt. **53**, 557-571 (2006).
[53] R. A. Campos, B. E. A. Saleh, and M. C. Teich, "Quantum-mechanical lossless beam splitter: SU(2) symmetry and photon statistics," Phys. Rev. A **40**, 1371-1384 (1989).
[54] D. M. Greenberger, M. A. Horne, and A. Zeilinger, in "Bell's Theorem, Quantum Theory and Conceptions of the Universe", edited by M. Kafatos (Springer Netherlands, Dordrecht, 1989), pp. 69–72.
[55] K. Hammerer, A. S. Sørensen, and E. S. Polzik, "Quantum interface between light and atomic ensembles," Rev. Mod. Phys. **82**, 1041-1093 (2010).
[56] C. H. Bennett, G. Brassard, C. Crépeau, R. Jozsa, A. Peres, and W. K. Wootters, "Teleporting an unknown quantum state via dual classical and Einstein-Podolsky-Rosen channels," Phys. Rev. Lett. **70**, 1895-1899 (1993).
[57] A. Furusawa, J. L. Sørensen, S. L. Braunstein, C. A. Fuchs, H. J. Kimble, and E. S. Polzik, "Unconditional Quantum Teleportation," Science **282**, 706-709 (1998).
[58] S. Takeda, T. Mizuta, M. Fuwa, P. van Loock, and A. Furusawa, "Deterministic quantum teleportation of photonic quantum bits by a hybrid technique," Nature **500**, 315-318 (2013).
[59] A. E. Ulanov, D. Sychev, A. A. Pushkina, I. A. Fedorov, and A. I. Lvovsky, "Quantum Teleportation Between Discrete and Continuous Encodings of an Optical Qubit," Phys. Rev. Lett. **118**, 160501 (2017).
[60] S. L. Braunstein, "Quantum error correction for communication with linear optics," Nature **394**, 47-49 (1998).
[61] M. Gu, C. Weedbrook, N. C. Menicucci, T. C. Ralph, and P. van Loock, "Quantum computing with continuous-variable clusters," Phys. Rev. A **79**, 062318 (2009).
[62] P. Jouguet, S. Kunz-Jacques, A. Leverrier, P. Grangier, and E. Diamanti, "Experimental demonstration of long-distance continuous-variable quantum key distribution," Nat. Photonics **7**, 378-381 (2013).
[63] B. Julsgaard, J. Sherson, J. I. Cirac, J. Fiurášek, and E. S. Polzik, "Experimental demonstration of quantum memory for light," Nature **432**, 482-486 (2004).
[64] J. Appel, E. Figueroa, D. Korystov, M. Lobino, and A. I. Lvovsky, "Quantum Memory for Squeezed Light," Phys. Rev. Lett. **100**, 093602 (2008).
[65] D. V. Vasilyev, I. V. Sokolov, and E. S. Polzik, "Quantum volume hologram," Phys. Rev. A **81**, 020302 (2010).
[66] K. Jensen, W. Wasilewski, H. Krauter, T. Fernholz, B. M. Nielsen, M. Owari, M. B. Plenio, A. Serafini, M. M. Wolf, and E. S. Polzik, "Quantum memory for entangled continuous-variable states," Nat. Physics **7**, 13-16 (2011).
[67] A. N. Vetlugin, I. V. Sokolov, "Addressable parallel cavity-based quantum memory," Eur. Phys. J. D **68**, 269 (2014).
[68] A. N. Vetlugin and I. V. Sokolov, "Multivariate quantum memory as controllable delayed multi-port beamsplitter," EPL **113**, 64005 (2016).
[69] A. S. Losev, T. Y. Golubeva, A. D. Manukhova, and Y. M. Golubev, "Two-photon bunching inside a quantum memory cell," Phys. Rev. A **102**, 042603 (2020).
[70] F. Lenzini, J. Janousek, O. Thearle, M. Villa, B. Haylock, S. Kasture, L. Cui, H.-P. Phan, D. V. Dao, H. Yonezawa, P. K. Lam, E. H. Huntington, and M. Lobino, "Integrated photonic platform for quantum information with continuous variables," Sci. Adv. **4**, eaat9331 (2018).

[71] F. Kaiser, B. Fedrici, A. Zavatta, V. D'Auria, and S. Tanzilli, "A fully guided-wave squeezing experiment for fiber quantum networks," Optica **3**, 362-365 (2016).
[72] G. Masada, K. Miyata, A. Politi, T. Hashimoto, J. L. O'Brien, and A. Furusawa, "Continuous-variable entanglement on a chip," Nat. Photonics **9**, 316-319 (2015).
[73] M. I. Kolobov, "The spatial behavior of nonclassical light," Rev. Mod. Phys. **71**, 1539-1589 (1999).
[74] R. Tualle-Brouri, A. Ourjoumtsev, A. Dantan, P. Grangier, M. Wubs, and A. S. Sørensen, "Multimode model for projective photon-counting measurements," Phys. Rev. A **80**, 013806 (2009).
[75] C. Kim and P. Kumar, "Quadrature-Squeezed Light Detection Using a Self-Generated Matched Local Oscillator," Phys. Rev. Lett. **73**, 1605-1608 (1994).
[76] I. V. Sokolov, "Schrödinger cat states in continuous variable non-Gaussian networks," Phys. Lett. A **384**, 126762 (2020).
[77] L.-M. Duan, G. Giedke, J. I. Cirac, and P. Zoller, "Inseparability Criterion for Continuous Variable Systems," Phys. Rev. Lett. **84**, 2722-2725 (2000).
[78] J. L. Everett, P. Vernaz-Gris, G. T. Campbell, A. D. Tranter, K. V. Paul, A. C. Leung, P. K. Lam, and B. C. Buchler, "Time-reversed and coherently enhanced memory: A single-mode quantum atom-optic memory without a cavity," Phys. Rev. A **98**, 063846 (2018).
[79] B. Julsgaard, A. Kozhekin, and E. S. Polzik, "Experimental long-lived entanglement of two macroscopic objects," Nature **413**, 400-403 (2001).
[80] Z. Y. Ou, S. F. Pereira, and H. J. Kimble, "Realization of the Einstein-Podolsky-Rosen paradox for continuous variables in nondegenerate parametric amplification," Appl. Phys. B **55**, 265-278 (1992).
[81] Z. Y. Ou, S. F. Pereira, H. J. Kimble, and K. C. Peng, "Realization of the Einstein-Podolsky-Rosen paradox for continuous variables," Phys. Rev. Lett. **68**, 3663-3666 (1992).
[82] D. B. Horoshko and M. I. Kolobov, "Generation of monocycle squeezed light in chirped quasi-phase-matched nonlinear crystals," Phys. Rev. A **95**, 033837 (2017).
[83] A. S. Sørensen and K. Mølmer, "Entangling atoms in bad cavities," Phys. Rev. A **66**, 022314 (2002).
[84] W. Wasilewski and M. G. Raymer, "Pairwise entanglement and readout of atomic-ensemble and optical wave-packet modes in traveling-wave Raman interactions," Phys. Rev. A **73**, 063816 (2006).
[85] N. I. Masalaeva, A. N. Vetlugin, and I. V. Sokolov, "Cavity-assisted squeezing and entanglement: non-adiabatic effects and optimal cavity-atomic ensemble matching," Phys. Scr. **95**, 034009 (2020).
[86] J. S. Neergaard-Nielsen, B. M. Nielsen, C. Hettich, K. Mølmer, and E. S. Polzik, "Generation of a Superposition of Odd Photon Number States for Quantum Information Networks," Phys. Rev. Lett. **97**, 083604 (2006).
[87] B. C. Sanders, "Entangled coherent states," Phys. Rev. A **45**, 6811-6815 (1992).
[88] M. Takeoka and M. Sasaki, "Conditional generation of an arbitrary superposition of coherent states," Phys. Rev. A **75**, 064302 (2007).
[89] A. Laghaout, J. S. Neergaard-Nielsen, I. Rigas, C. Kragh, A. Tipsmark, and U. L. Andersen, "Amplification of realistic Schrödinger-cat-state-like states by homodyne heralding," Phys. Rev. A **87**, 043826 (2013).
[90] D. V. Sychev, A. E. Ulanov, A. A. Pushkina, M. W. Richards, I. A. Fedorov, and A. I. Lvovsky, "Enlargement of optical Schrödinger's cat states," Nat. Photonics **11**, 379-382 (2017).
[91] I. Afek, O. Ambar, and Y. Silberberg, "High-NOON States by Mixing Quantum and Classical Light," Science **328**, 879-881 (2010).
[92] F. Shafiei, P. Srinivasan, and Z. Y. Ou, "Generation of three-photon entangled state by quantum interference between a coherent state and parametric down-conversion," Phys. Rev. A **70**, 043803 (2004).
[93] Y. Israel, L. Cohen, X.-B. Song, J. Joo, H. S. Eisenberg, and Y. Silberberg, "Entangled coherent states created by mixing squeezed vacuum and coherent light," Optica **6**, 753-757 (2019).
[94] https://doi.org/10.21979/N9/UFSSOT.